\documentclass[12pt,preprint]{aastex}

\newcommand{\OM}      {${\Omega}_{\rm M}$}
\newcommand{\OL}      {${\Omega}_{\Lambda}$}
\newcommand{\OK}      {${\Omega}_{\rm K}$}
\newcommand{\OT}      {${\Omega}_{\rm total}$}

\newcommand{\OMLflat} {${\Omega}_{\rm M}+{\Omega}_{\Lambda}=1$}

\newcommand{\Dm}      {${\Delta{m_{15}}}$}
\newcommand{\lcdm}    {$\Lambda{\rm CDM}$}
\newcommand{\scriptM} {$\cal{M}$}
\newcommand{\chisqnu} {$\chi^2/{\rm DoF}$}
\newcommand{\chisq}   {$\chi^2$}

\newcommand{\snia}    {SN~Ia}
\newcommand{\snib}    {SN~Ib}

\newcommand{\sneia}   {SNe~Ia}

\newcommand{\wgloszsys}{$0.13$}

\newcommand{\wglosz}             {$-1.05^{+0.13}_{-0.12}~{\rm (stat}~1\sigma{)} \pm 0.13~{\rm (sys)}$}

\newcommand{\wgloszjoint}        {$-1.07^{+0.09}_{-0.09}~{\rm (stat}~1\sigma{)} \pm 0.13~{\rm (sys)}$}

\newcommand{\omglosz}            {$0.274^{+0.033}_{-0.020}~{\rm (stat}~1\sigma{)}$}

\newcommand{\omgloszjoint}       {$0.267^{+0.028}_{-0.018}~{\rm (stat}~1\sigma{)}$}

\newcommand{\glosznumsnejoint}{162}

\newcommand{\saltnumsnejoint}{178}

\newcommand{\chisqnugloszlcdm}          {$0.96$}
\newcommand{\chisqnuglosz}              {$0.96$}
\newcommand{\chisqnugloszjointlcdm}     {$0.90$}
\newcommand{\chisqnugloszjoint}         {$0.91$}

\newcommand{\stddevgloszlcdm}          {$0.20$}

\newcommand{\stddevgloszjointlcdm}     {$0.23$}

\newcommand{\chisqnusaltjointlcdm}{$2.76$}

\newcommand{\stddevsaltjointlcdm}{$0.28$}

\newcommand{\sigmadisp}{$\sigma_{\rm add}$}

\newcommand{\numWsne}          {60} 
 
\newcommand{\numnearbysne}     {45}

\usepackage{graphicx}
\usepackage{longtable}
\graphicspath{{Figures/}}

\makeatletter

\usepackage{natbib}

\makeatother
\begin{document}

\title{Observational Constraints on the Nature of Dark Energy: First Cosmological Results from the ESSENCE Supernova Survey}

\shorttitle{Dark Energy from the ESSENCE Survey}

\slugcomment{Submitted to ApJ on November 21, 2006}

\author{
{W.~M.~Wood-Vasey}\altaffilmark{1}, 
{G.~Miknaitis}\altaffilmark{2}, 
{C.~W.~Stubbs}\altaffilmark{1,3},
{S.~Jha}\altaffilmark{4,5}, 
{A.~G.~Riess}\altaffilmark{6,7},
{P.~M.~Garnavich}\altaffilmark{8}, 
{R.~P.~Kirshner}\altaffilmark{1},
{C.~Aguilera}\altaffilmark{9}, 
{A.~C.~Becker}\altaffilmark{10}, 
{J.~W.~Blackman}\altaffilmark{11}, 
{S.~Blondin}\altaffilmark{1}, 
{P.~Challis}\altaffilmark{1}, 
{A.~Clocchiatti}\altaffilmark{12}, 
{A.~Conley}\altaffilmark{13}, 
{R.~Covarrubias}\altaffilmark{10}, 
{T.~M.~Davis}\altaffilmark{14}, 
{A.~V.~Filippenko}\altaffilmark{4}, 
{R.~J.~Foley}\altaffilmark{4}, 
{A.~Garg}\altaffilmark{1,3}, 
{M.~Hicken}\altaffilmark{1,3}, 
{K.~Krisciunas}\altaffilmark{8,16},
{B.~Leibundgut}\altaffilmark{17}, 
{W.~Li}\altaffilmark{4}, 
{T.~Matheson}\altaffilmark{18}, 
{A.~Miceli}\altaffilmark{10}, 
{G.~Narayan}\altaffilmark{1,3},
{G.~Pignata}\altaffilmark{12}, 
{J.~L.~Prieto}\altaffilmark{19}, 
{A.~Rest}\altaffilmark{9}, 
{M.~E.~Salvo}\altaffilmark{11}, 
{B.~P.~Schmidt}\altaffilmark{11}, 
{R.~C.~Smith}\altaffilmark{9}, 
{J.~Sollerman}\altaffilmark{14, 15}, 
{J.~Spyromilio}\altaffilmark{17}, 
{J.~L.~Tonry}\altaffilmark{20}, 
{N.~B.~Suntzeff}\altaffilmark{9, 16}, and
{A.~Zenteno}\altaffilmark{9}
}
\email{wmwood-vasey@cfa.harvard.edu}

\nopagebreak

\altaffiltext{1}{Harvard-Smithsonian Center for Astrophysics, 60 Garden Street, Cambridge, MA 02138}
\altaffiltext{2}{Fermilab, P.O. Box 500, Batavia, IL 60510-0500}
\altaffiltext{3}{Department of Physics, Harvard University, 17 Oxford Street, Cambridge, MA 02138}
\altaffiltext{4}{Department of Astronomy, 601 Campbell Hall, University of California, Berkeley, CA 94720-3411}
\altaffiltext{5}{Kavli Institute for Particle Astrophysics and Cosmology, Stanford Linear Accelerator Center, 2575 Sand Hill Road, MS 29, Menlo Park, CA 94025}
\altaffiltext{6}{Space Telescope Science Institute, 3700 San Martin Drive, Baltimore, MD 21218}
\altaffiltext{7}{Johns Hopkins University, 3400 North Charles Street, Baltimore, MD 21218}
\altaffiltext{8}{Department of Physics, University of Notre Dame, 225 Nieuwland Science Hall, Notre Dame, IN 46556-5670}
\altaffiltext{9}{Cerro Tololo Inter-American Observatory, Casilla 603, La Serena, Chile}
\altaffiltext{10}{Department of Astronomy, University of Washington, Box 351580, Seattle, WA 98195-1580}
\altaffiltext{11}{The Research School of Astronomy and Astrophysics, The Australian National University, Mount Stromlo and Siding Spring Observatories, via Cotter Road, Weston Creek, PO 2611, Australia}
\altaffiltext{12}{Pontificia Universidad Cat\'olica de Chile, Departamento de Astronom\'ia y Astrof\'isica, Casilla 306, Santiago 22, Chile}
\altaffiltext{13}{Department of Astronomy and Astrophysics, University of Toronto, 50 Saint George Street, Toronto, ON M5S 3H4, Canada}
\altaffiltext{14}{Dark Cosmology Centre, Niels Bohr Institute, University of Copenhagen, Juliane Maries Vej 30, DK-2100 Copenhagen \O, Denmark}
\altaffiltext{15}{Department of Astronomy, Stockholm University, AlbaNova, 10691 Stockholm, Sweden}
\altaffiltext{16}{Department of Physics, Texas A\&M University, College Station, TX 77843-4242}
\altaffiltext{17}{European Southern Observatory, Karl-Schwarzschild-Strasse 2, D-85748 Garching, Germany}
\altaffiltext{18}{National Optical Astronomy Observatory, 950 North Cherry Avenue, Tucson, AZ 85719-4933}
\altaffiltext{19}{Department of Astronomy, Ohio State University, 4055 McPherson Laboratory, 140 West 18th Avenue, Columbus, OH 43210}
\altaffiltext{20}{Institute for Astronomy, University of Hawaii, 2680 Woodlawn Drive, Honolulu, HI 96822}

\pagebreak
\begin{abstract}
We present constraints on the dark energy equation-of-state parameter, 
$w=P/(\rho c^2)$,
using \numWsne\ Type Ia supernovae (\sneia) from the ESSENCE supernova survey.
We derive a set of constraints on the nature of the dark
energy assuming a flat Universe.
By including constraints on (\OM, $w$) from 
baryon acoustic oscillations, we obtain a value for a static
equation-of-state parameter 
$w=$\wglosz\ and
\OM=\omglosz\ with a best-fit \chisqnu\ of \chisqnuglosz. 
These results are consistent with those reported by the SuperNova
Legacy Survey in a similar program measuring supernova distances and
redshifts.
We evaluate sources of systematic error that afflict supernova
observations and present Monte Carlo simulations that explore these
effects.
Currently, the largest systematic currently with the potential to affect our measurements is the
treatment of extinction due to dust in the supernova host galaxies.
Combining our set of ESSENCE \sneia\ with the SuperNova Legacy Survey \sneia, we obtain a joint constraint of
$w=$\wgloszjoint, \OM=\omgloszjoint\
with a best-fit \chisqnu\ of \chisqnugloszjoint.
The current \sneia\ data are fully consistent with a cosmological constant.
\end{abstract}


\pagebreak

\section{Introduction: Supernovae and Cosmology}
\label{sec:introduction}

We report the analysis of \numWsne\ Type Ia supernovae (\sneia) discovered in the
course of the ESSENCE program (Equation of State: Supernovae trace
Cosmic Expansion---an NOAO Survey Program) from 2002 to 2005.  
The aim of
ESSENCE is to measure the history of cosmic expansion over the past 
5~billion years with sufficient precision to distinguish whether the
dark energy is different from a cosmological constant at the $\sigma_w=\pm0.1$ 
level.  
Here we present our first results and show that we are well on our way towards that goal.
Our present data
are fully consistent with a $w=-1$, flat Universe, 
and our uncertainty in $w$, the parameter that
describes the cosmic equation of state, analyzed in the way we outline
here, will shrink below $0.1$ for models of constant $w$ as the ESSENCE program is completed.
Other approaches to using the luminosity distances have been suggested
to constrain possible cosmological models.  
We here provide the ESSENCE observations in a convenient form suitable for a
testing a variety of models.\footnote{\url{http://www.ctio.noao.edu/essence/}}

As reported in a companion paper \citep{miknaitis07}, ESSENCE is
based on a supernova search carried out with the 4-m
Blanco Telescope at the Cerro Tololo Inter-American Observatory (CTIO) 
with the prime-focus MOSAIC II 64 Megapixel CCD camera.  
Our search produces densely sampled $R$-band and $I$-band 
light curves for supernovae in our fields.  As described in that
paper, we optimized the search to provide the best constraints on $w$,
given fixed observing time and the properties of the MOSAICII camera
and CTIO 4-m telescope.
Spectra from a variety of large telescopes, including Keck, VLT,
Gemini, and Magellan, allow us to determine supernova types and
redshifts. We have paid particular attention to the central problems of
calibration and systematic errors that, when the survey is complete
in 2008, will be more important to the final precision of our
cosmological inferences than statistical sampling errors for about 200
objects.

This first cosmological report from the ESSENCE survey 
derives some properties of dark energy from the
sample presently in hand, which is still small enough that the
statistics of the sample size make a noticeable contribution to the
uncertainty in dark-energy properties.  
But our goal is to set out the systematic uncertainties in a clear way so
that these are exposed to view and so that we can concentrate our
efforts where they will have the most significant effect.
To infer luminosity distances
to the ESSENCE supernovae over the redshift interval $0.15$--$0.70$, we
employ the relations developed for \sneia\ at low redshift~\citep{jha06c} 
among their light-curve shapes, colors, and intrinsic luminosities.
The expansion history from $z\approx0.7$ to the
present provides leverage to constrain the equation-of-state parameter for the
dark energy as described below.  
In \S\ref{sec:introduction} we sketch the context of the ESSENCE program.
In \S\ref{sec:distances} we show from a set of
simulated light curves that this particular implementation of 
light-curve analysis is consistent, with the same cosmology emerging from
the analysis as was used to construct the samples, and that the
statistical uncertainty we ascribe to the inference of the dark-energy
properties is also correctly measured.  This modeling of our analysis
chain gives us confidence that
the analysis of the actual data set is reliable and its uncertainty is
correctly estimated.
Section~\ref{sec:systematics} delineates the systematic errors we confront, estimates
their present size, and indicates some areas where improvement can be
achieved.
Section~\ref{sec:cosmology} describes the sample and provides the estimates of dark
energy properties using the ESSENCE sample.
The conclusions of this work are given in \S\ref{sec:conclusions}.

\subsection{Context}

Supernovae have been central to cosmological measurements from the
very beginning of observational cosmology.  
\citet{shapley19} employed supernovae against the ``island universe'' hypothesis arguing that
objects such as SN~1885A in Andromeda would have $M=-16$ which was ``out of
the question.''
Edwin Hubble \citep{hubble29b} noted ``a
mysterious class of exceptional novae which attain luminosities that
are respectable fractions of the total luminosities of the systems in
which they appear.''  These extra-bright novae were dubbed ``supernovae''
by \citet{baade34} and divided into two classes, based on
their spectra, by \citet{minkowski41}. Type~I supernovae (SNe~I) have
no hydrogen lines while Type~II supernovae (SNe~II) show H$\alpha$ and
other hydrogen lines.

The high luminosity and observed homogeneity of the first handful of
SN~I light curves prompted \citet{wilson39} to suggest that they be
employed for fundamental cosmological measurements, starting with time
dilation of their characteristic rise and fall to distinguish true
cosmic expansion from ``tired light.''  
After the \snib\ subclass was separated from the \sneia\ \citep[see][for a review]{filippenko97} this line of investigation has
grown more fruitful as techniques of photometry have improved and
as the redshift range over which supernovae have been well
observed and confirmed to have standard light-curve shapes and luminosities has increased \citep{rust74,leibundgut96,riess97,goldhaber01,riess04b,foley05,hook05,conley06,blondin06}.
Within the uncertainties, the results agree with the predictions of
cosmic expansion and provide a fundamental test that the underlying
assumption of an expanding universe is correct.

Evidence for the homogeneity of \sneia\ comes from their small scatter in the Hubble diagram.
 \citet{kowal68} compiled data
for the first well-populated Hubble diagram of SNe~I. The
1$\sigma$ scatter about the Hubble line was $0.6$~mag, but Kowal
presciently speculated that supernova distances to individual objects
might eventually be known to 5-10\% and suggested that ``[i]t may even be possible
to determine the second-order term in the redshift-magnitude relation
when light curves become available for very distant supernovae.''  
Precise
distances to \sneia\ enable tests for the linearity of the Hubble
law and provide evidence for local deviations from the local Hubble
flow, attributed to density inhomogeneities in the local universe
\citep{riess95,riess97,zehavi98,bonacic00,radburn-smith04,jha06c}.
While \snia{} cosmology is not dependent on the value of $H_0$, 
it is sensitive to deviations from a homogeneous Hubble flow and
these regional velocity fields may limit our ability to estimate
properties of dark energy, as emphasized by \citet{hui06}
and by \citet{cooray06}.  Whether the best strategy is to map
the velocity inhomogeneities thoroughly or to skip over them by using
a more distant low-redshift sample remains to be demonstrated.
We have used a lower limit of redshift $z>0.015$ in constructing our sample of
\sneia{}.

The utility of \sneia\ as distance
indicators results from the demonstration that the intrinsic brightness of each \snia\ is closely
connected to the shape of its light curve.
As the sample of well-observed \sneia\ grew, some
distinctly bright and faint objects were found. For
example, SN~1991T \citep{filippenko92a,phillips92} and SN~1991bg \citep{filippenko92b,leibundgut93} were of different luminosity, and their light curves were
not the same, either.  The possible correlation of the shapes of
supernova light curves with their luminosities had been explored by
\citet{pskovskii77b}.  More homogeneous photometry from CCD
detectors, more extreme examples from larger samples, and more reliable distance estimators enabled \citet{phillips93} to establish
the empirical relation between light-curve shapes and supernova
luminosities. The Cal\'an-Tololo sample
\citep{hamuy96} and the CfA sample \citep{riess99,jha06a}, have been used to improve the methods for using supernova
light curves to measure supernova distances.  Many variations on
Phillips' idea have been developed, including
\Dm\ \citep{phillips99}, 
MLCS \citep{riess96,jha06c}, 
DM15 \citep{prieto06}, 
stretch \citep{goldhaber01}, 
CMAGIC \citep{wang03}, 
and SALT \citep{guy05}.

These methods are capable of achieving the 10\% precision for
supernova distances that \cite{kowal68} foresaw 40 years ago.
In the ESSENCE analysis, we have used a version of the
\citet{jha06c} method called MLCS2k2.  We have compared it with the results
of the SALT \citep{guy05} light-curve fitter used by the SuperNova Legacy Survey \citep[SNLS;][]{astier06}.
This comparison provides a test: if the
two approaches do not agree when applied to the same data they cannot
both be correct.  As shown in \S\ref{sec:distances}, SALT and this version of
MLCS2K2, with our preferred extinction prior, are in excellent accord when applied to the same data.  
While gratifying, this agreement does not prove they are both correct.
Moreover, as described in \S\ref{sec:cosmology}, the cosmological
results depend somewhat on the assumptions about SN host-galaxy
extinction that are employed.
This has been an ongoing problem in supernova cosmology.
The work of \citet{lira95} demonstrated the empirical fact that although SN Ia have a
range of colors at maximum light, they appear to reach the same
intrinsic color about 30--90 days past maximum light, independent of
light curve shape.

\citet{riess96} used de-reddened \snia\ data to show that near maximum
light intrinsic color differences existed with fainter \sneia
appearing redder than brighter objects and then used this information
to construct an absorption-free Hubble diagram.  Given a good set of
observations in several bands, the reddening for individual supernovae
can then be determined and the general relations between supernova
luminosity and the light curve shapes in many bands
can be established \citep{hamuy96,riess99,phillips99}.
The initial detections of cosmic acceleration employed 
either these individual absorption corrections \citep{riess98} 
or a full-sample statistical absorption correction \citep{perlmutter99}.
Finding the best approach to
this problem, whether by shifting observations to the infrared,
limiting the sample to low-extinction cases, or making other
restrictive cuts on the data, is an important area for future work.
Some ways to explore this issue are sketched in \S\ref{sec:cosmology}.

\citet{kowal68} recognized that second-order terms in cosmic expansion
might be measured with supernovae once the precision and redshift
range grew sufficiently large.  
More direct approaches with the {\it Hubble Space Telescope (HST)} were imagined
by \citet{colgate79} and with special clarity by \citet{tammann79}.  
Tammann anticipated that {\it HST} photometry of \sneia\ at $z\approx0.5$ would lead
to a
direct determination of cosmic deceleration and that the time dilation
of \snia\ light curves would be a fundamental test of the expansion
hypothesis.  While {\it HST} languished on the ground after the Challenger
disaster, this line of research was attempted from the ground at
the European Southern Observatory (ESO) by a Danish group in 1986--1988.  
Their cyclic CCD imaging of the search fields used image registration, convolution and subtraction, and real-time data analysis \citep{hansen87}.  
Alas, the rate of \sneia\ in their fields was lower
than they had anticipated, and only one \snia, SN~1988U was
discovered and monitored in two years of effort
\citep{hansen87,norgaard-nielsen89}.  More effective searches by the
Lawrence Berkeley Lab (LBL) group exploiting larger CCD detectors
and sophisticated detection software showed that this approach could be 
made practical and used to find
significant numbers of high-redshift \sneia~\citep{perlmutter95}.

By 1995, two groups, the LBL-based Supernova Cosmology Project (SCP)
and the High-Z Supernova Search Team \citep[HZT;][]{schmidt98}) were
working in this field.
The first \snia\ cosmology results using 7 high-redshift \sneia
\citep{perlmutter97} found a Universe consistent with \OM$=1$ but
subsequent work by the SCP \citep{perlmutter98} 
and by the HZT \citep{garnavich98} revised this initial finding
to favor a lower value of \OM.
At the January 1998 meeting of the American
Astronomical Society both teams reported that the \snia\ results favored a universe that would expand without limit, but at that time neither team claimed the
Universe was accelerating.  
The subsequent publication of
stronger results based on larger samples by the HZT \citep{riess98} and by the SCP \citep{perlmutter99} provided a big
surprise.  The supernova data showed
that \sneia\ at $z\approx0.5$ were about $0.2$~mag dimmer than expected in an open
universe and pointed firmly at an accelerating universe 
\citep[for first-hand accounts, see][]{overbye99,riess00b,filippenko01b,kirshner02,perlmutter03};
reviews are given by \citet{leibundgut01}, \citet{filippenko05a} and others.

The supernova route to cosmological understanding continues to
improve.  One source of uncertainty has been the small sample of very
well observed low-redshift supernovae~\citep{hamuy96,riess99}.  The most recent contribution
is the summary of CfA data obtained in 1997--2001 \citep{jha06a}, but
significantly enhanced samples from the CfA~\citep{hicken06} 
together with 
new data from the Katzman Automatic Imaging Telescope \citep[KAIT;][]{li00,filippenko01a,filippenko05b}, 
from the Carnegie SN Program \citep{hamuy06},
from the Supernova Factory \citep{wood-vasey04,copin06},
and from the Sloan Digital Sky Survey II Supernova Survey \citep[SDSS~II;][]{frieman04,dilday05} are in prospect.  
As the low-$z$ sample approaches
200~objects, the size of the sample will cease to be a source of
statistical uncertainty for the determination of cosmological
parameters.  
As described in \S\ref{sec:systematics}, systematic errors of calibration and
K-correction will ultimately impose the limits to understanding dark energy's properties, and we
are actively working to improve these areas \citep{stubbs06}.

Some of the potential sources of systematic error in the high-$z$ sample have been examined.
The fundamental assumption is that distant \sneia\ can be analyzed using
the methods developed for the low-$z$ sample.  Since
nearby samples show that the \sneia\ in elliptical galaxies have a different
distribution in luminosity than the \sneia\ in spirals \citep{hamuy00,gallagher05,neill06,sullivan06b}, morphological classification of the
distant sample may provide some useful clues to help improve the
cosmological inferences \citep{williams03}. For example, \citet{sullivan03}
 showed that restricting the SCP sample to \sneia\ in
elliptical galaxies gave identical cosmological results to the complete sample,
which is principally in spiral galaxies.  The possibility of grey dust
raised by \citet{aguirre99a,aguirre99b} was examined by \citet{riess00a} and  and by \citet{nobili05} through infrared observations of high-$z$ supernovae and was put to rest by
the very high-redshift observations of \citet{riess04b}. 
Improved methods for handling the vexing problems of absorption by
dust have been developed by \citet{knop03} and by \citet{jha06c}.
These questions are described in more detail in \S\ref{sec:extinction}. 

The question of whether distant supernovae have spectra that are the
same as nearby supernovae has been investigated by 
\citet{coil00}, \citet{lidman05}, \citet{matheson05}, \citet{hook05}, \citet{howell05}, and \citet{blondin06}.
The more telling question of whether these spectra
evolve in the same way as those of nearby objects was approached by \citet{foley05}.
In all cases, the evidence points toward nearby
supernovae behaving in the same way as distant ones, bolstering
confidence in the initial results.  This observed consistency does not mean that the
samples are identical, only that the variations between the nearby
and distant samples are successfully accounted for by the methods
currently in hand.  We do not know whether this will continue to be the case as
future investigations press for more stringent limits on cosmological
parameters \citep{albrecht06}.

The highest redshift \snia\ data~\citep{riess04b} show the qualitative
signature expected from a mixed dark-energy/dark-matter cosmology.
That is, they show cosmic
deceleration due to dark matter preceded the current era of cosmic
acceleration due to dark energy.
The sign of
the observed effect on supernova apparent magnitudes reverses---\sneia\ at $z\sim 0.5$ appear $0.2$~mag dimmer than expected
in a coasting cosmology but the very distant supernovae whose light comes
from $z>1$ appear brighter than they would in that cosmology.  By
itself, this turnover is a very encouraging sign that supernova cosmology does
not founder on grey dust or even on a simple evolution of supernova
properties with cosmic epoch.  As part of this analysis, \citet{riess04b} 
constructed the ``gold'' sample of high-$z$ and low-$z$ supernovae
whose observations met reasonable criteria for inclusion in an
analysis of all of the published light curves and spectra using a
uniform method of deriving distances from the light curves.

The analysis of the gold sample provided an estimate of the
time derivative of the equation-of-state parameter, $w$, for dark energy.  
These
observations are very important conceptually because the simplest fact
about the cosmological constant as a candidate for dark energy is
that it should be constant (i.e., $w' = dw/dz = 0$).  The observations are
consistent with a constant dark energy over the redshift range out to
$z\approx 1.6$. Other forms of dark energy could satisfy the observed
constraints, but this observational test is one that the cosmological
constant could have failed.  In the analysis of the ESSENCE data
presented in \S\ref{sec:cosmology}, we use the supernova data to constrain the
properties of $w$, as first carried out by \citet{white98} and by \citet{garnavich98}.  
This parameterization of dark energy by $w$ is not the only possible approach.  
A more detailed approach is to compare the observational data to a specific model and, for example, try to reconstruct the dark energy
scalar-field potential~\cite[see, for example,][]{li06}.
A more agnostic view is that we are simply measuring the expansion
history of the universe, and a kinematic description of that history
in terms of expansion rate, acceleration, and jerk
\citep{riess04b,rapetti06} covers the facts without assuming the nature of dark energy.

The ESSENCE project was conceived to tighten the constraints on dark
energy at $z\approx0.5$ to reveal any discrepancy between the
observations and the leading candidate for dark energy, the
cosmological constant.  A simple way to express this is that we aim for 
a 10\% uncertainty in the value of $w$.  This program is similar to the approach
of the SuperNova Legacy Survey (SNLS) being carried out at the
Canada-France-Hawaii Telescope, and we compare our methods and results
to theirs \citep{guy05,astier06} at several points in the analysis below.

The SNLS has taken the admirable step of
publishing their light curves online and making the code of their light-curve
fitting program, SALT, available for public inspection and use\footnote{\url{http://snls.in2p3.fr/conf/release/}}.
Making the light curves public, as was done
for the results of the HZT and its successors \citet{riess98,tonry03,barris04,krisciunas05,clocchiatti06}, by \citet{knop03}, by \citet{riess04b} for the very high
redshift {\it HST} supernova program, and for the low-$z$ data of \citet{hamuy96}, \citet{riess99}, 
and \citet{jha06a}, provides the opportunity for others to
perform their own analysis of the results. In addition to exploring a
variety of approaches to analyzing our own \snia\ observations, we show
the first joint constraints from ESSENCE and SNLS, and some joint
constraints derived from combining these with the
\citet{riess04b} gold sample in \S\ref{sec:cosmology}.

\pagebreak
\section{Luminosity Distance Determination}
\label{sec:distances}

The physical quantities of interest in our cosmological measurements
are the redshifts and distances to a set of space-time points in the
Universe.  The redshifts come from spectra and the luminosity
distances, $D_L$, come from the observed flux of the supernova combined with
our understanding of \snia\ light curves from nearby objects.

Extracting a luminosity distance to a supernova from observations of its light
curve necessitates a number of assumptions.  
We use the observations of nearby supernovae to establish the relations
between color, light-curve shape in multiple bands, and peak luminosity. These nearby 
observations attain high signal-to-noise ratios, and the nearby objects can be observed
in more passbands (including infrared) than faint distant objects. We assume
that  the resulting method of converting  light curves to luminosity distances 
applies at all redshifts. The observed spectral uniformity of supernovae over a range of redshift~\citep{lidman05,hook05,blondin06}
supports this approach. 
We assume that $R_V$, the ratio of selective to absolute extinction, is 
independent of redshift. 
Below in \S\ref{sec:extinction}, 
we test the potential systematic effect of departures from this assumption.
We adopt an astrophysically sensible prior distribution of host-galaxy extinction
properties, with a redshift dependence that is derived from the simulations we present below. 

Our approach is to conduct comprehensive
simulations of the ESSENCE data and analysis.
As described by \citet{miknaitis07}, we use this same approach to
explore our photometric performance.
For the aspects of our analysis that are ``downstream'' of the
light-curve generation, we generate sets of synthetic light curves and
subject them to our analysis pipeline.
In this way we can test the performance of our distance-fitting tools,
and by exaggerating various systematic errors (zeropoint offsets, etc.) we
can assess the impact of these effects on our determination of $w$. 

We must recognize and emphasize that in the era of precision \snia
cosmology (constraining dark energy properties, rather than just
detecting its existence), careful attention to systematic errors is of
paramount importance: shifts of a few hundredths of a magnitude can
lead to constraints on $w$ that change by $0.1$. Different, yet
defensible, choices in the analysis chain may show such effects.

\subsection{Extracting Luminosity Distances from Light Curves: Distance Fitters}

We use the MLCS2k2 method of \citet{jha06c}
as the primary tool to
derive relative luminosity distances to our \sneia.
For comparison, we also provide the results obtained using the 
Spectral Adaptive Lightcurve Template (SALT)
fitter of \citet{guy05} on the ESSENCE light curves.
SALT was used in the recent cosmological results paper from the SNLS~\citep[][, hereafter A06]{astier06}.  
We provide a consistent and comprehensive set of distances obtained to
nearby, ESSENCE, and SNLS supernovae for each luminosity-distance
fitting technique.
The ESSENCE light curves used in this analysis were presented by 
\citet{miknaitis07} and we provide them online,
together with our set of previously published light curves for nearby \sneia, 
for the convenience of those interested.\footnote{\url{http://www.ctio.noao.edu/essence/}}
Additional \snia\ light-curve fitting methods will be further explored in future ESSENCE analyses.
Understanding the behavior of our distance determination method is
critical to our goal of quantifying the uncertainties of our analysis chain.

MLCS2k2 and SALT, as well as the light-curve ``stretch'' approach
used by \citet{perlmutter97,perlmutter99}, \citet{goldhaber01} and \citet{knop03}, exploit the fact that
the rate of decline, the color, and the intrinsic luminosity of \sneia
are correlated.
At present we treat \sneia\ as a single-parameter family, 
and the distance fitting techniques use multi-color light
curves to deduce a luminosity distance and host-galaxy reddening for
each supernova.
Previous papers have shown 
that the different techniques produce relative luminosity
distances that scatter by $\sim 0.10$~mag for an individual
\snia~\citep[e.g.,][]{tonry03}, but this scatter is uncorrelated with redshift.
As a consequence, the cosmology results are insensitive to the distance fitting technique.
However, as described by \citet{miknaitis07}, the measurement of the
equation-of-state parameter hinges on subtle distortions in the Hubble
diagram, so we have undertaken a comprehensive set of simulations
to understand potential biases introduced by MLCS2k2.

The MLCS2k2 approach \citep{riess96,riess98,jha06c} to determining luminosity
distances uses well-observed nearby \sneia\ to establish a set of
light-curve templates in multiple passbands.
The parameters $\Delta$ (roughly equivalent to the variation in peak visual luminosity, this parameter characterizes intrinsic color, rate
of decline, and peak brightness), $A_V$ (the $V$-band extinction of the
supernova light in its host galaxy), and $\mu$ (the
distance modulus) are then determined by fitting each multi-band set of
distant supernova light curves to redshifted versions of these
templates.
Jha et al. (2006c) present results from MLCS2k2 based on nearby SN
Ia. Here we have modified MLCS2k2 for application to both high and
low-redshift \sneia. We begin with a rest-frame model of the \snia\ in its
host galaxy, and then propagate the model light curves through the
host-galaxy extinction, K-correction, Milky Way extinction to the
detector, incorporating the measured passband response (including the
atmosphere for ground-based observations). We then fit this model
directly to the natural-system observations. This forward-modeling
approach has particular advantages in application to the more sparsely
sampled (in color and time) data typical of high-redshift SN searches.

The SALT method of
\citet{guy05}, which was used for the SNLS first-results analysis of
A06, constructs a fiducial \snia\ template using combined
spectral and photometric information, then transforms this template
into the rest frame of the \snia, and finally calculates a flux, stretch, and
generalized color.  The color parameter in SALT is notable in that
it includes both the intrinsic variation in \snia\ color
and the extinction from dust in the host galaxy within a single parameter
(in contrast, MLCS2k2 attempts to separate these components of the observed colors for each supernova). 
While the reddening vector (attenuation vs. color excess) 
is similar to the \snia\ color vs. absolute magnitude relation, 
the two sources of correlated color and luminosity variation are not identical.

The stretch and color parameters of SALT were used by A06 to
estimate luminosity distances by fitting for the stretch-luminosity
and color-luminosity relationships in the nearby sample and 
applying those to the full \sneia\ sample.
Given that the SALT color parameter conflates the two physically distinct 
phenomena of host-galaxy extinction and \snia\ color variation, 
it is remarkable and perhaps
a source of deep insight that this treatment works as well as it does.
Because of both survey selection effects and possible demographic
shifts in the host environments of \sneia
we would not expect that the proportion of reddening from dust and
from intrinsic variation would remain constant with redshift as this
approach assumes.
However, the SALT/A06 method does seem to work quite well in practice. 

\subsection{Sensitivity to Assumptions about the Host-Galaxy Extinction Distribution: Extinction Priors}
\label{sec:extinction_prior}

The best way to treat host-galaxy extinction is a serious question for
this work and for the field of supernova cosmology.  The Bayesian approach we
use is detailed in \S\ref{sec:priors}.  Here we describe simulations that are
designed to evaluate the effects of those methods.

There have been four basic approaches to combining reddening
measurements with astrophysical knowledge to determine the host galaxy
extinction along the line-of sight:
(1)  assume that linear $A_V$ is the natural space for extinction and assume a flat prior~\citep{perlmutter99,knop03};
(2)  use models of the dust distribution in galaxies~\citep{hatano98,commins04,riello05} to model line-of-sight extinction values~\citep{riess98,tonry03,riess04b};
(3)  assume that the distribution of host-galaxy $A_V$ follows an exponential form~\citep{jha06c}, based on observed distributions of $A_V$ in nearby \sneia;
and
(4)  self-calibrate within a set of low-$z$ \sneia\ to obtain a consistent color+$A_V$ relationship and assume that relation for the full set~\citep{astier06}.

Approach (1) assumes the least prior knowledge
about the distribution of $A_V$ and produces a Gaussian
probability distribution for the fitted luminosity distance.  
However, this approach weakens the ability to separate intrinsic \snia
color from $A_V$ and results in a fit parameter $A_V$ that is a mixture
of the two.
An $A_V$ that is truly
related to the dust extinction should never be negative.
The probability prior with $-\infty < A_V < +\infty$ is not the 
natural range over which to assume a flat distribution.
The physically reasonable prior on $A_V$ should be strictly positive.
One approach is to base the prior for absorption on the distribution of dust in galaxies.
Theoretical modeling of dust distributions in galaxies, such as that
of \citet{hatano98}, \citet{commins04}, and \citet{riello05}, provides a 
physically motivated dust distribution.
This method represents approach (2) above and is the method we adopt here.
In contrast, \citet{jha06c} empirically derived an exponential $A_V$
distribution from MLCS2k2 fits to nearby \sneia\ by assuming a particular
color distribution of \sneia.
This distribution was derived using the empirical fact that \sneia\ 
reach a common color about 40 days past maximum light~\citep{lira95}.
They found an exponential distribution of $A_V$, 
\begin{equation}
p(A_V) \propto \exp \left(\frac{-A_V}{\tau}\right),
\label{eq:default}
\end{equation}
where $\tau=0.46$~mag.
Unfortunately, the highest-extinction objects drive the
tail of this exponential and significantly affect the fit,
resulting in a prior sensitive to sample selection, which differs
significantly in high-redshift searches compared to the nearby objects
studied by \citet{jha06c}.

A06 analyzed the results of the SALT \snia\ light-curve fitter with
approach (4) and have systematic sensitivities that are similar to
those of approach (1).

We use MLCS2k2 as our main analysis tool.  
We designate approach (1) the ``flatnegav'' prior and approach (3)
the ``default'' prior and discuss both of these further in
\S\ref{sec:priors}.
Approach (2) is based on a galactic line-of-sight or ``glos'' prior on $A_V$:
\begin{equation}
\hat{p}(A_V) \propto \frac{A}{\tau} \exp\left({\frac{-A_V}{\tau}}\right) + 
\frac{2B}{\sqrt{2\pi}\sigma} \exp\left(\frac{-A_V^2}{2\sigma^2}\right),
\label{eq:glos}
\end{equation}
where $A=1$, $B=0.5$, $\tau=0.4$, $\sigma=0.1$, and $\hat{p}(A_V) \equiv 0$ for $A_V < 0$. 
This exponential plus one-sided narrow Gaussian ``glos'' prior is based on the host-galaxy dust models of 
\citet{hatano98}, \citet{commins04}, and \citet{riello05}.  
As described below, we have modeled our selection effects with
redshift to adapt the ``glos'' prior into the ``glosz'' prior that is the basis for our
analysis.  We feel this approach leverages our best understanding of the effects of extinction and selection.

\begin{figure}
\plottwo{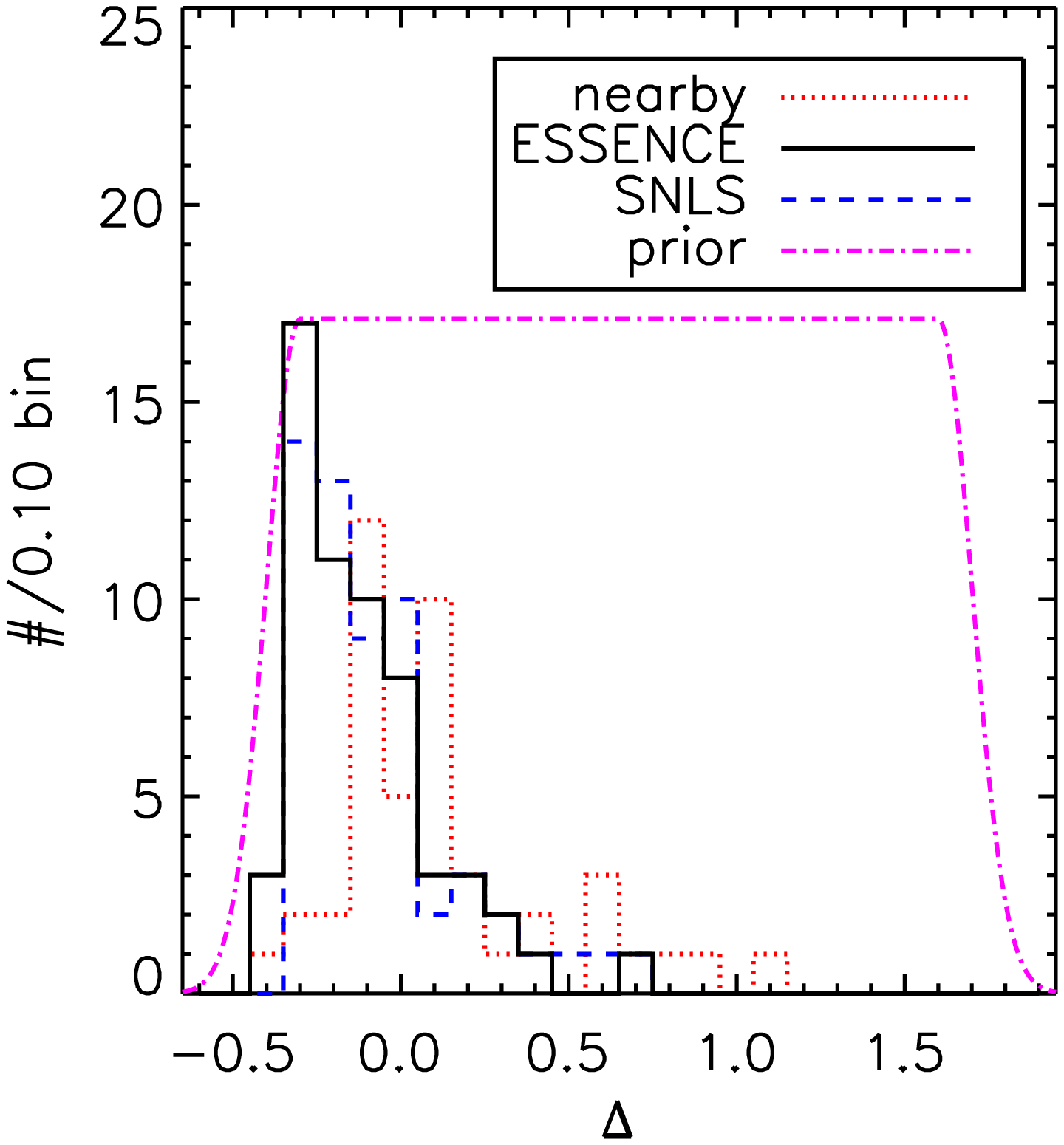}{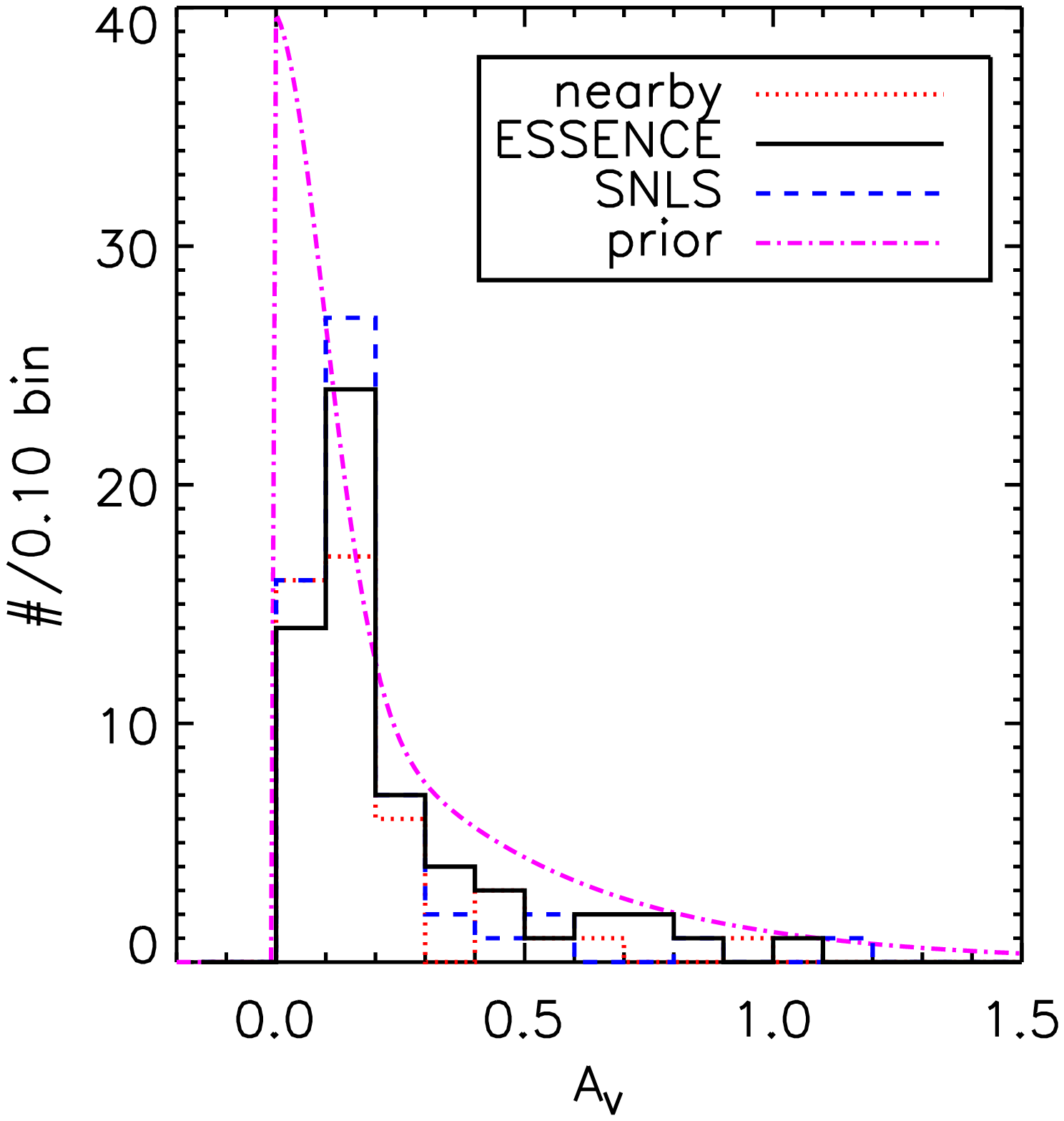}
\caption{
The distribution of the MLCS2k2 light-curve width parameter $\Delta$
and $A_V$ for the MLCS2k2 fits with the ``glosz'' prior 
to the nearby (dotted line), ESSENCE (solid line), and SNLS (dashed line) \sneia\ considered in this paper.  
The ``glosz'' prior (dotted-dashed line) is shown here for $z=0$
where it is equivalent to the ``glos'' prior. 
Note that we are mixing two slightly different things in showing the
prior with these estimated mean fit parameters.
The prior, which directly relates to the mode, is not expected to
match the a posteriori mean distribution of the fit parameters.
See Fig.~\ref{fig:cut} for the ESSENCE selection effect as a function of redshift.
See Table~\ref{tab:mlcs_fits_prior_glosz} for the full set of MLCS2k2 light-curve fit results for these \sneia.
}
\label{fig:mlcs_fits_prior_glosz}
\end{figure}

\begin{figure}
\plottwo{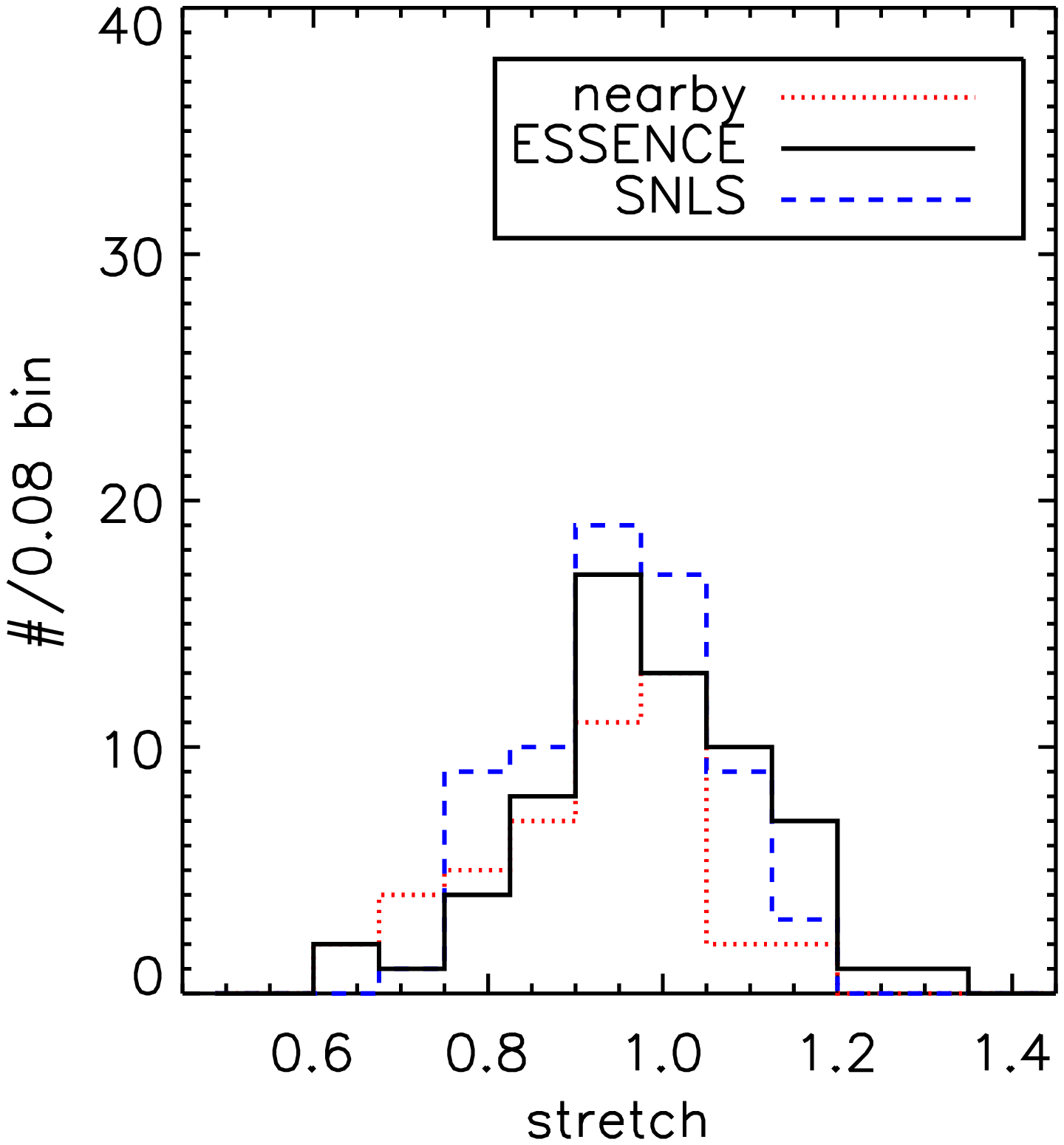}{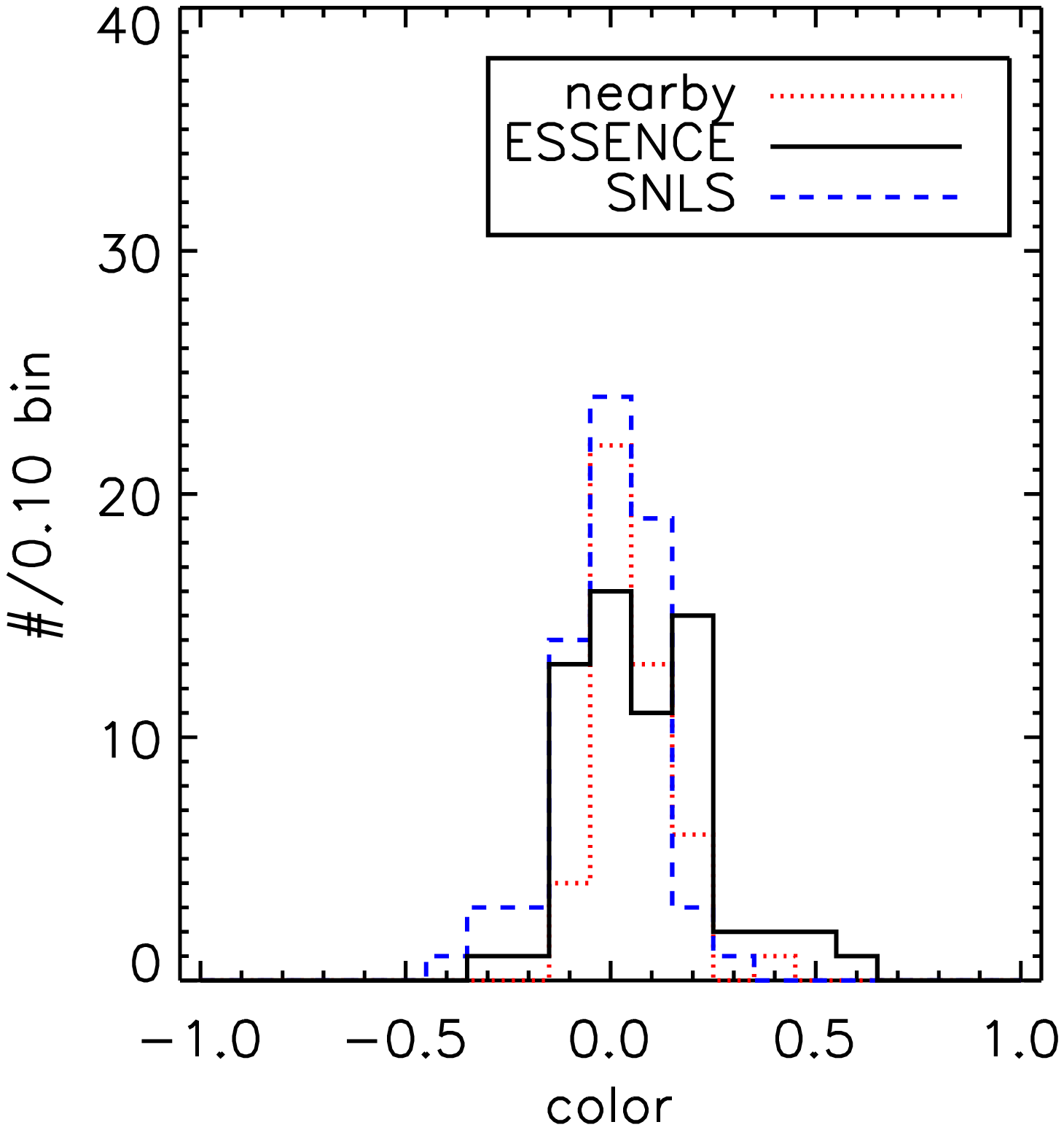}
\caption{The distribution of the SALT light-curve stretch and the estimated
color plus extinction 
for the nearby (dotted line), ESSENCE (solid line), and SNLS (dashed line) \sneia\ considered in this paper.  
The priors for SALT are effectively flat for stretch and color, 
and SALT quotes minimum \chisq\ values instead of 
the estimated mean parameter values of MLCS2k2.
See Table~\ref{tab:salt_fits} for the full set of SALT light-curve fit results for these \sneia.
}
\label{fig:salt_fits}
\end{figure}

Figs.~\ref{fig:mlcs_fits_prior_glosz} and \ref{fig:salt_fits} show the
distribution of the fit parameters and overlay the prior distribution
assumed for each of these approaches.
Fig.~\ref{fig:mlcs_vs_salt} compares the fit distances and
extinction/color parameters of the MLCS2k2 ``glosz'' and SALT fit
results for the ESSENCE, SNLS, and nearby samples.
The distribution of recovered $\Delta$ and $A_V$ match their imposed
priors for MLCS2k2 ``glosz'' while the stretch and color fit
parameters from SALT show a consistent distribution for the three
different sets of \sneia.

\subsection{ESSENCE Selection Effects and the Motivation for a Redshift-Dependent 
Extinction  Prior}
\label{sec:selection}

We examined the effect of the survey
selection function on the expected demographics of the
ESSENCE \sneia\ and explored the interplay between extinction, Malmquist bias, 
and our observed light curves.
To determine the impact of the selection bias, we developed
a Monte Carlo simulation of the ESSENCE search.  We created a range of
supernova light curves that match the properties of the nearby sample,
added noise based on statistics from actual ESSENCE photometry, and then fit the
resulting light curves in the same way the real events are analyzed.
In this way we estimated the impact of subtle biases, although 
this simulation cannot test for errors in our light-curve model
or population drift with redshift.

Based on its low-redshift training set, MLCS2K2 is able
to output a finely sampled light curve
given a redshift ($z$), distance modulus ($\mu$), light-curve shape parameter ($\Delta$),
host extinction ($A_V$), host extinction law ($R_V$), date of rest-frame
$B$-band maximum light ($t_0$), Milky Way reddening ($E(B-V)_\mathrm{MW}$), and the
bandpasses of the observations.
At a given redshift we calculated a distance modulus, $\mu_\mathrm{true}$, from
the luminosity distance for the standard cosmology ($\Omega_m=0.3$,
$\Omega_\Lambda =0.7$) and that distance modulus plus an 
assumed $M_B=-19.5$ for \sneia\ set the brightness for our simulated
supernovae. Varying the assumed cosmology does not significantly impact 
the simulation results since we are comparing the input distance modulus
with the recovered distance modulus, $\mu_\mathrm{obs}$, which is independent of
the cosmology. 

At each of a series of fixed redshifts, we created $\sim1000$ simulated light curves with
parameters chosen from random distributions.
The light-curve width, $\Delta$, was selected from the \citet{jha06c} distribution measured
from the low-$z$ sample. The $\Delta$ distribution is approximately
a Gaussian peaking at $\Delta =-0.15$ with an extended tail out to
$\Delta =1.5$.
The host extinction for each simulated event, $A_V$, was
selected from either the \citet{jha06c}
distribution (``default'') estimated from the local sample or from a
``galaxy line-of-sight'' estimation (``glos''). The ``default''
distribution was an exponential decay with index $0.46$~mag and set to
zero for $A_V < 0.0$~mag. The ``glos'' distribution is also set to zero for
$A_V < 0.0$~mag and combines a narrow Gaussian with a exponential tail for
$A_V > 0.0$~mag (see Eq.~\ref{eq:glos}). The extinction law is assumed to be $R_V = 3.1$.
The Milky Way reddening [$E(B-V)_\mathrm{MW}$] distribution was constructed from
the \citet{schlegel98} (hereafter SFD) reddening maps that cover the ESSENCE fields. The $E(B-V)_\mathrm{MW}$ was
measured for 10,000 random locations in each ESSENCE field and the
reddening was selected from the sum of the histograms (see Figure~2).
The dates of observation for a simulated \snia\ were
based on the actual dates of ESSENCE 4-m observations. An
ESSENCE field was chosen at random from the list of monitored
fields and a date of maximum, $t_0$,
selected to fall randomly between the Modified Julian Date (MJD) of
the first and last observation of an observing season. The
simulated light curve was then interpolated for only those
dates that ESSENCE took images. With each ESSENCE field observation,
we estimated the magnitude in $R$ and $I$ that provided a 10$\sigma$
photometric detection based on the seeing and sky brightness. The
signal-to-noise ratio (SNR) for each simulated light-curve point was then
scaled from the 10$\sigma$\ detection magnitude, assuming the noise
was dominated by the sky background.

For each date of ESSENCE observation, we have a simulated noiseless
magnitude and an estimate of the SNR of the observation.
To each simulated observation
we added an appropriate random value in flux space 
selected from a normal distribution with a width
corresponding to the predicted SNR. 

MLCS2k2 was then used to fit the simulated light curves
and provide estimates of $\mu$, $\Delta$, $A_V$, and $t_0$,
assuming a fixed $R_V = 3.1$. MLCS2k2 required an initial
guess of the date of maximum, an estimate achieved by selecting
from a normal distribution about the true date with a
$1\sigma$\ width of 2 days. The SFD Milky Way reddening
was also required in MLCS2k2 and was provided from the
true reddening after adding an uncertainty of 10\%.
Finally, in the real ESSENCE data we discarded
supernovae when the MLCS2k2 reduced $\chi^2$ indicated
a very poor fit. For the simulated light curves, we
dropped events from the sample if the reduced $\chi^2$ exceeded
2.

\subsubsection{Deriving an Extinction Prior from the Simulation Results}

\begin{figure}
\plotone{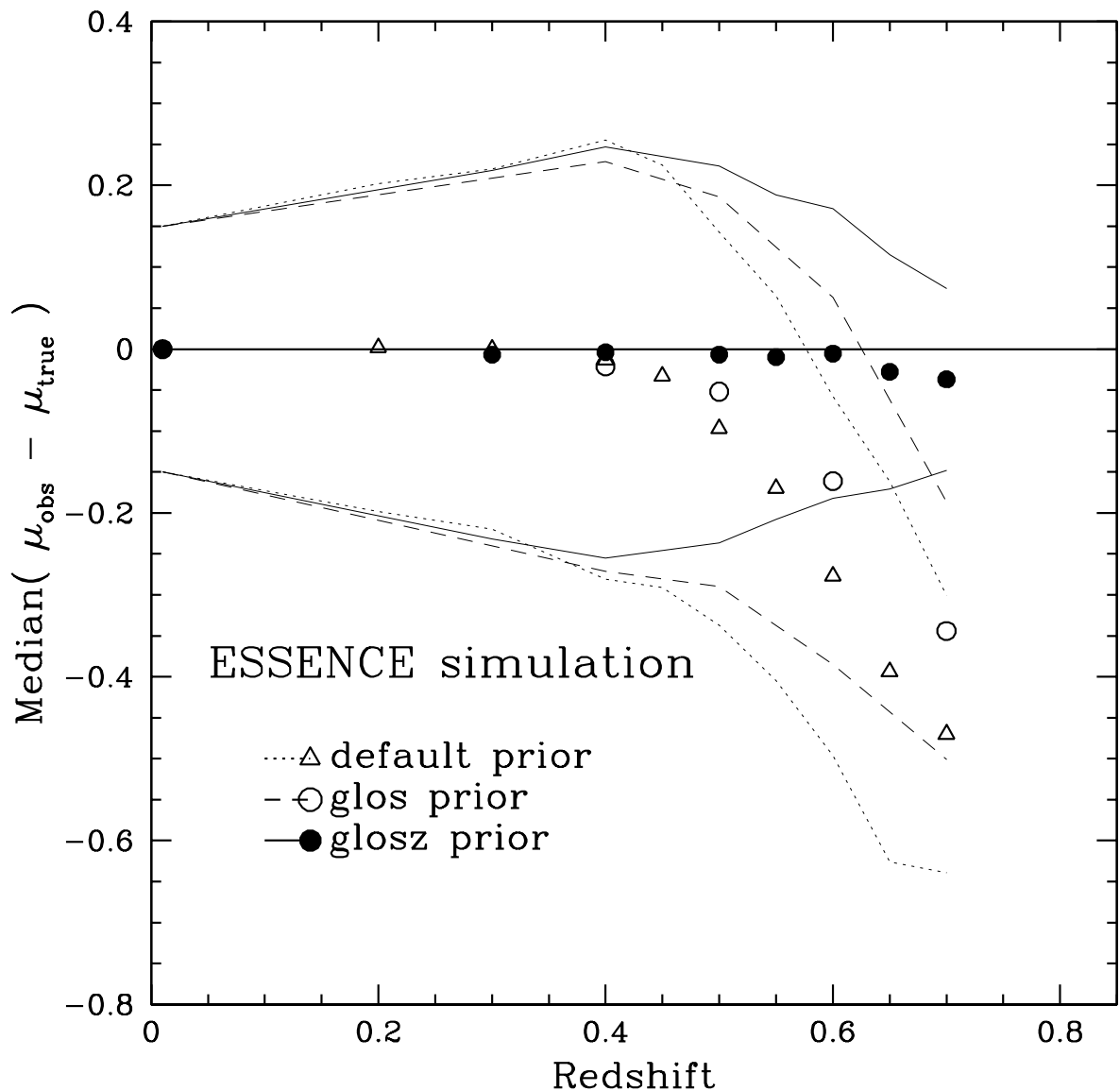}
\caption{
The median of the distance modulus error as a function of redshift for
the simulated data sets.
The points show the median value of the difference between the input
$\mu_{true}$ and recovered $\mu_{obs}$ of about 1000 simulated
supernovae at each redshift.
The lines indicate the root-mean-square spread of the recovered distance modulus.
}
\label{fig:simulatepriors}
\end{figure}

Simulated ESSENCE samples were created at a range of redshifts
out to $z=0.70$ and the light curves that passed the detection
criteria from the actual ESSENCE search were fit with MLCS2k2. 
The fitting was done with the ``default'' prior and the ``glos'' prior (with
corresponding $A_V$ distributions). The difference between the
``true'' (input) distance modulus and recovered (fit) distance
modulus, $\Delta\mu$, was calculated for each event and the mean, median, and
dispersion for the ensemble were calculated at each redshift.
The median $\Delta\mu$ of the simulations was within $0.03$~mag
for $z < 0.45$, but at higher redshift the simulated supernovae
were estimated to be brighter than the input supernovae by more than
$0.2$~mag. This bias results from the loss of faint events (large
$A_V$ and large $\Delta$) from the sample as the distance increases.
In a sense, this is a classic Malmquist bias, but here it is
caused by an uninformed prior.  These results are shown in Fig.~\ref{fig:simulatepriors}.

\begin{figure}
\plotone{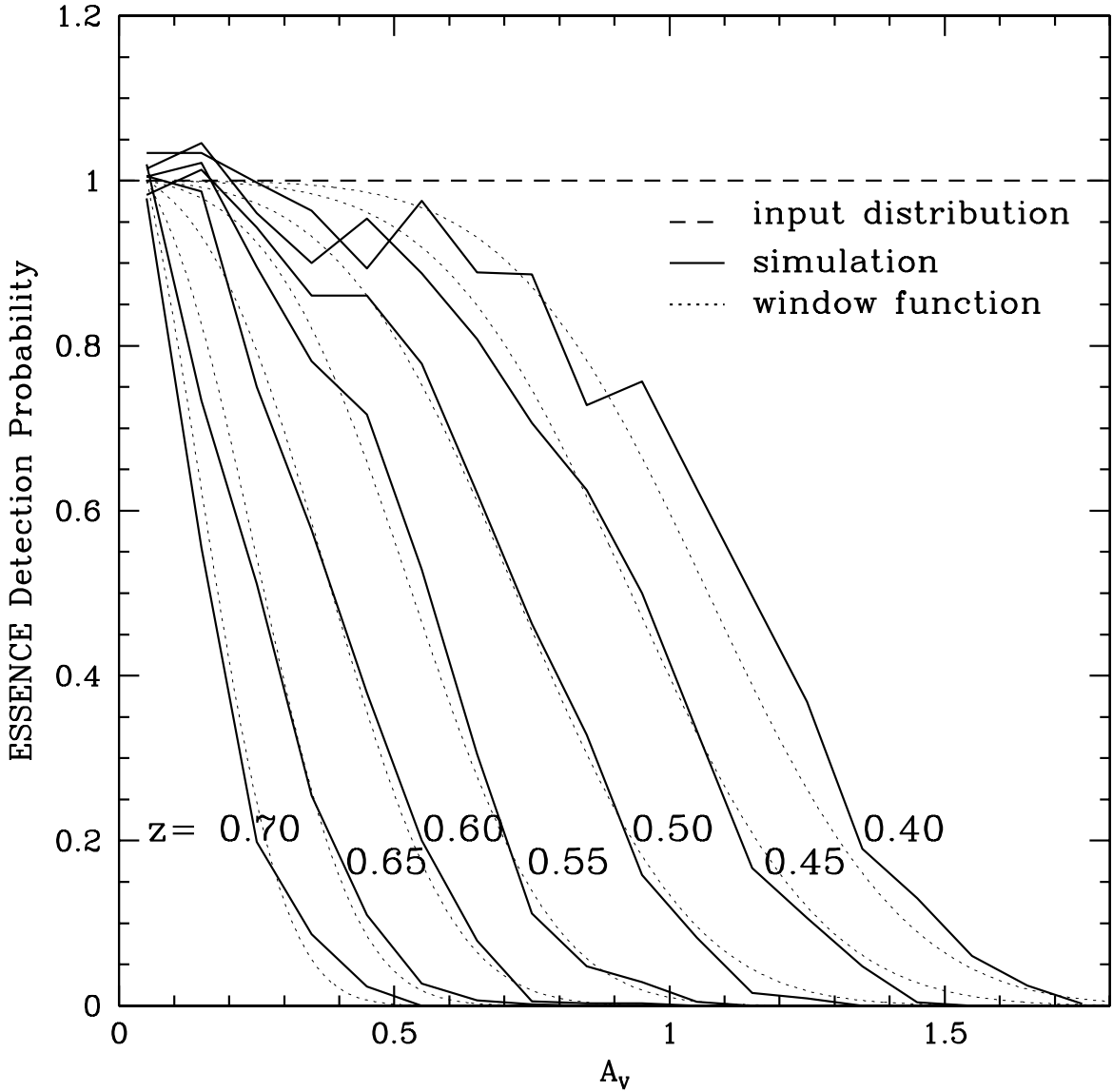}
\caption{
The recovered distribution of visual extinctions for simulated
supernovae in the ESSENCE sample if the input distribution were
uniform in $A_V$ out to large extinctions.
The curves are fit to determine the parameters of the window function
(see Table~\ref{tab:gloszwindow}) which is then used to modify the ``glos'' 
prior a function of redshift into the ``glosz'' prior.
We estimate the SNLS selection function as extending $+0.2$ in
redshift deeper than the ESSENCE selection function.  
}
\label{fig:cut}
\end{figure}

The decreasing ability to observe large $A_V$ events as the redshift increases
(see Fig.~\ref{fig:cut}) 
makes it clear that a using single $A_V$ prior for all redshifts is not correct. 
Because
events with large $A_V$ and large $\Delta$ are lost at large
redshift due to the magnitude limits of the search, 
we should adjust the prior as a function of $z$ to account
for these predictable losses. 
Applying redshift-dependent window functions to the basic ``glos''
prior provides a much better prior as a function of redshift.

We fit the recovered $A_V$ distributions derived from the
simulations, which start with a uniform $A_V$, to a
window function based on the error function (integral of a Gaussian), 
and two parameters describe where that function drops to
half its peak value ($A_{1/2}$) and the width of the transition
($\sigma_A$). The window function $W$ has the form
\begin{equation}
W(A_V,A_{1/2},\sigma_A) = 1-
{1\over{\sqrt{\pi}}}\int_{-\infty}^{(A_V-A_{1/2})\sigma_A}
e^{-x^2}\;dx
\label{eq:gloszwindow}
\end{equation}
where $A_{1/2}$ and $\sigma_A$ are functions of $z$
and estimated from the simulations. A similar process was
applied to the $\Delta$ distribution and a table providing
the parameters is given in Table~\ref{tab:gloszwindow}.
We embody this prescription in the ``glosz'' prior we use for our main MLCS2k2 light-curve fitting.
The ``glosz'' prior is the ``glos'' prior modified by the window
functions in $A_V$ and $\Delta$. The simulations using
the ``glosz'' prior provide a median $\Delta\mu$ within
0.03 mag for $z < 0.7$, which we judge to be satisfactory performance.

\subsection{Comparison of MLCS2k2 and SALT Luminosity Distance Fitters}

The release of the source code to the SALT
fitter~\citep{guy05} 
makes a modern \snia\ light-curve fitter
fully accessible and available to the community.
This public release of SALT allows us to compare the results of our MLCS2k2
distance fitter with the SALT fitter used in the SNLS first results
paper~\citep{astier06}.
We present the results of SALT fits to our nearby and ESSENCE samples
in Table~\ref{tab:salt_fits}.  To compute the distance moduli we
quote in that table, we assume the $\alpha=1.52$, $\beta=1.57$ values
from A06.
To calibrate the additional dispersion to add to the distance moduli
of MLCS2k2 and SALT, we fit a \lcdm\ model to the nearby sample alone
and derived the additional \sigmadisp\ to added in quadrature to recover
\chisqnu$=1$ for the nearby sample.  This \sigmadisp\ is related to the 
intrinsic dispersion of the absolute luminosity of \sneia, but is not
precisely the same both because the light-curve fitters include 
varying degrees of model uncertainty and because the light curves
of the \sneia\ are subject to photometric uncertainty.
We find \sigmadisp$=0.10$ for MLCS2k2 with the ``glosz'' prior and
\sigmadisp$=0.13$ for SALT.  These values should be added to the 
$\sigma_\mu$ uncertainties given Tables~\ref{tab:mlcs_fits_prior_glosz} and \ref{tab:salt_fits}
Fig.~\ref{fig:mlcs_vs_salt} visually demonstrates that
the relative luminosity distances using the SALT 
light-curve fitter agree, within uncertainties, with the MLCS2k2 distances when the latter are fit
using the ``glosz'' $A_V$ prior.

\subsection{Testing the Recovery of Cosmological Models Using Simulations of the ESSENCE Dataset}
\label{sec:simulation}

In order to assess the reliability with which we recover cosmological 
parameters, we have simulated 100 sets 
of 100 light curves representing both the nearby
and the ESSENCE light curves.
Table~\ref{tab:mlcs_cuts} presents the quality cuts for MLCS2k2 we derived
from these simulated light-curve sets.
Our light-curve goodness-of-fit cuts, when applied to these simulated light curves
(see Table~\ref{tab:mlcs_cuts}) and combined with the same external constraints of baryon acoustic oscillations \citep[BAO;][]{eisenstein05} and flatness, allow us to recover our input cosmology of
$(\Omega_M=0.3, \Omega_\Lambda=0.7, w=-1)$ to within $\pm0.11$ in $w$.
This $\pm0.11$ uncertainty on an individual measurement of
$w$ is matched by the $\sigma=0.11$ distribution of recovered $w$
values from the 100 sets of simulated light curves.
This confirms our statistical error estimate on $w$; the estimated
uncertainty matches the distribution, and within the self-consistent
realm of synthetic and analyzed light curves based on MLCS2k2 our
estimates of luminosity distance are not biased.

\pagebreak

\section{Potential Sources of Systematic Error}
\label{sec:systematics}

Here we identify and assess sources of systematic error
that could afflict our measurements. These can be divided into two groups.
Certain sources of systematic error may introduce perturbations
either to individual photometric data points or to the distances or
redshifts estimated to the \sneia.  Others affect the
data in a more or less random fashion and produce excess
{\em scatter} in the Hubble diagram. 
Errors that are uncorrelated with either distance or redshift will not
bias the cosmological result.
These sources of photometric error are detailed by \citet{miknaitis07};
we summarize those results here in Table~\ref{tab:photscatter}.
We add these effects in quadrature to the statistical uncertainties
given by the luminosity distance fitting codes 
for each \snia\ distance measurement: $\sigma_\mu^{\rm phot scatter}=0.026$~mag.

In \S\ref{sec:distances} we discussed our testing of the
MLCS2k2 fitter on simulated data sets that replicate the data quality
of the ESSENCE and nearby \sneia.
We explore the issue of host-galaxy extinction further in \S\ref{sec:extinction} \& \ref{sec:priors}.
The interaction of Malmquist bias and selection effects 
with the extinction and color distribution of \sneia
is discussed in \S\ref{sec:malmquist}.

Any non-cosmological difference in measurements
of nearby and distant \sneia\ has the potential to perturb our
measurement of $w$. Table~\ref{tab:systematics} lists potential
systematic effects of this sort. We present both our estimate of the sensitivity
$dw/dx$ of the equation-of-state parameter to each potential systematic
effect and our best estimate of the potential size of
the perturbation, ${\Delta}x$. The upper bound on the bias introduced
in $w$ is then ${\Delta}w = dw/dx\times{\Delta}x$. 
\citet{miknaitis07} discusses the systematic uncertainties on $\mu$,
which we convert here to systematic uncertainties on $w$, due to
photometric errors from astrometric uncertainty on faint objects (${\Delta}w = 0.005$),
potential biases from the difference imaging (${\Delta}w = 0.001$),
and linearity of the MOSAIC~II CCD (${\Delta}w = 0.005$).
None of these contributed noticeably to the systematic uncertainty in
our measurement of $w$.
The rest of this section
describes how we appraised our additional potential sources of systematic
uncertainty.

The conclusion of this section is that 
our current overall estimate for the 1$\sigma$ equivalent systematic
uncertainty in a static equation-of-state parameter is 
${\Delta}w =$ \wgloszsys\ for our ``glosz'' analysis.

\subsection{Photometric Zeropoints}
\label{sec:zeropoint}

Supernova cosmology fundamentally depends on the ability to accurately
measure fluxes of objects over a range in redshift. Errors in photometric
calibration translate to errors in cosmology in two basic ways. 

Nearby objects at redshifts $<0.1$ play a crucial role in establishing
a comparison reference for cosmological measurements. 
ESSENCE is inefficient at finding and observing low-redshift objects
with the same telescope and detector system, 
so we use photometry of low-redshift \sneia\ in the literature
from our own work and that of others \citep[for the full list see][]{jha06c}. 
Using these external \sneia\ requires understanding the photometric calibration of
our
high-redshift sample relative to this low-redshift sample. 
Every supernova cosmology result to date has made use
of more or less the same low-redshift photometry, so any inaccuracies
in the nearby sample are a source of common systematic error for all \snia\ cosmology
experiments. Calibration of photometry at the $\sim1\%$~level required
to make precise inferences about the nature of dark energy is notoriously
difficult~\citep{stubbs06}.

Photometric miscalibration can result in a second, more insidious systematic error 
if there is an error in the relative flux scaling between the broad-band passbands.
This offset would distort the observed colors for the entire sample. 
Since these colors are used to infer
the extinction, even small color
errors result in significant biases in the measured distances.
After all, the inferred host galaxy extinction, $A_{V}$, 
is related to the measured color excess, $E(B-V)$, 
by $A_{V}\approx3.1 E(B-V)$ (for Milky Way-like dust). 
A color error in rest-frame
$B-V$ (observer-frame $R$, $I$ for ESSENCE) of $0.01$~mag can result
in $0.03$~mag error in extinction, an inaccuracy that would lead directly to a $3\%$
error in the distance modulus, or a $1.5\%$ error in the
distance.  We currently estimate our color zeropoint uncertainty at $0.02$~mag and 
our absolute zeropoint (relative to the nearby \sneia) uncertainty to be $0.02$~mag.
These respectively translate to $0.04$ and $0.02$ shifts in $w$ (see Table~\ref{tab:systematics}).

\citet{miknaitis07} describe the calibration program we undertook to measure the transmission of
the CTIO~4-m MOSAIC~II system with the $R$ and $I$ filters of the ESSENCE survey.
The calibration of the ESSENCE survey fields will be further
improved by an intensive calibration program we are undertaking on the
CTIO~4-m in 2006.  Together with the improved calibration
of the SDSS Southern Stripe by the SDSSII project, which overlaps
25\% of our ESSENCE fields, we aim to achieve 1\% photometric
calibration of our CTIO~4-m MOSAIC~II BVRI natural system.

We here use MLCS2k2 v004 with the \citet{bohlin04} values for the
magnitudes of Vega: i.e., \verb'alpha_lyr_stis_002.fits' with $R_{\rm
Vega}=0.033$~mag.  This value for $R_{\rm Vega}$ comes from
\citet{bessell98} but has been shifted down by $0.004$~mag as
\citet{bohlin04} suggest (from their $V_{\rm Vega}=0.026$~mag compared
to \citet{bessell98} $V_{\rm Vega}=0.030$~mag).

\subsection{K-Corrections and Bandpass Uncertainty}
\label{sec:kcorrection}

Uncertainty in the transmission function, typically called the bandpass,
of the optical path of the telescope+detector is an important and potentially
systematic effect. In this context, bandpass refers to the wavelength-dependent
throughput of the entire optical path, including atmospheric transmission,
mirror reflectivity, filter function, and CCD response. 
Since an error in the assumed bandpasses translates into a
redshift-dependent error in the supernova flux, it is important to
account for possible errors in the bandpass estimates.

The \emph{relative} error due to bandpass miscalibration is small
for objects with similar spectra, such as \sneia. Bandpass shape
errors are largely accounted for by the filter zeropoint calibration,
with residual errors corresponding to the difference between the spectral
energy distribution of the objects of interest and those of the calibration
sources. In the case of \snia\ observations, any residual zeropoint
error is absorbed when we marginalize over the ``nuisance parameter,''
\scriptM$=M_B - 5\log_{10} (H_0) + 25$~\citep{kim04}. 
This relative comparison results in a very small systematic error in
the cosmological parameters from a global calibration error across bandpasses. Moreover, variations in
atmospheric transmission are expected to contribute only random uncertainty.

However, the bandpass uncertainty becomes important when we compare
\sneia\ at different redshifts for which the bandpass samples
different spectral regions. In order to compare \sneia\ at multiple
redshifts, we need to perform a K-correction~\citep{leibundgut90,hamuy93a,kim96,nugent02}. 
That is, we assume a
spectral distribution for the supernova and convert the observed magnitude
to what it would have been had the supernova been at another redshift.
This process involves performing synthetic photometry of the assumed spectral
distribution over the assumed bandpass. We address the issue
of systematics arising from errors in the assumed spectral distribution
in the supernova evolution section, \S\ref{sec:snevolution}.
Here we address systematics arising from errors in our determination of the CTIO~4-m MOSAIC~II $R$ and $I$  bandpass functions.
Systematic effects on supernova cosmology that result from bandpass uncertainties are discussed more thoroughly by \citet{davis06}.

\label{sec:bandpass}

Calculating the effect of bandpass uncertainty is fairly difficult 
because of the arbitrary nature of the shape changes that might affect
the bandpass. However, we can make several general calculations. As
a first step, we take standard bandpasses and add white noise to represent
a miscalibrated filter. White noise contributes power on all scales,
so this approach adds small-scale discrepancies as well as large-scale warps or shifts
in the filter. By averaging over many such miscalibrated filters,
we can estimate the effect of filter miscalibration. Fig.~16 of \citet{davis06}
shows photometric error as a function of noise amplitude. A noise
amplitude of $0.02$ produces a typical deviation of $2$\% from the nominal
filter shape at any wavelength. Calibrating the bandpass to better
than $3\%$ allows us to keep the K-correction error introduced from a
mismeasurement of our effective bandpass to less than 0.005 mag 
(0.5\% in flux) and a systematic uncertainty of ${\Delta}w = 0.005$.

\subsection{Extinction}
\label{sec:extinction}

The most significant cause of variation in luminosity of \sneia\ is the
extinction experienced by the light from the \snia
due to scattering and absorption from dust in the host galaxy.  

Dust introduces a wavelength-depended diminution of a supernova's
light.  In the case of Milky Way dust, we correct for its effects by
using tabulated values as a function of Galactic longitude and
latitude measured by other means \citep[SFD][]{schlegel98}, being sure in our MLCS2k2 fits to properly account for its uncertainty and correlation across all observations.  For
dust in the supernova's host galaxy, we infer the extinction from the
reddening of each supernova's light curve.

However, the slope of differential reddening, characterized in
the \citet{cardelli89} extinction model by the parameter $R_V$, may vary.
The nominal value of $R_V$ for the Milky
Way is $3.1$, but different lines of sight within
our galaxy have values of $R_V$ that vary from $2.1$ to $5.1$.  
Studies of $R_V$ in other galaxies have been more
limited because we lack sources of known color and luminosity
with which to probe the dust.

Because we use the supernova rest-frame $B-V$ color to determine the reddening of each \snia,
and the distance modulus to a supernova is corrected by a value approximately
three times the inferred reddening, extinction correction 
magnifies any source of systematic error in a supernova's observed
effective color. Systematic color errors can result from photometry errors,
redshift-dependent K-correction errors, and evolution in the colors of supernovae.

Using the IR-emission maps of the Galaxy from the all-sky COBE/DIRBE
and IRAS/ISSA maps, SFD have estimated the dust column density around
the sky, which can then be translated to a color excess. This analysis
has largely superseded the work of \citet{burstein78}, who used radio
HI measurements and a relationship between gas and extinction to
estimate the color excess across the sky. \citet{burstein03} has
reanalyzed the IR and HI measurements and finds that Milky Way extinctions
are more precisely derived using the IR method.
However, \citet{burstein03} still finds a
discrepant value for extinction at the poles, with SFD providing extinctions
that are $E(B-V)=0.02$~mag higher than what the HI measurements indicate. 
\citet{burstein03} suggests as a possible explanation for the discrepancy
that SFD may predict too large an extinction in areas with high gas-to-dust ratios. 
\citet{finkbeiner99} precisely estimated their sensitivities to these 
systematics and concluded they had controlled them to $0.01$~mag.
The ESSENCE program targets fields at high Galactic
latitude to minimize Galactic extinction. Although nearby and distant \sneia
are both affected by the assumed Milky Way extinction, the nearby
objects are observed in $B-V$, whereas the $z\approx0.5$ objects
are observed in $R-I$. An $E(B-V)=0.02$ difference in extinction at the pole
leads to approximately a $0.02$ mag difference in the relative distances
between $z=0$ and $z=0.5$ objects, assuming a Galactic reddening
law, host-galaxy corrections based on rest-frame $B-V$ color, and distances
based on $V$.
For this analysis, we use the SFD extinction map values with an uncertainty of 16\% for each individual \snia\ 
but assume an additional $0.01$~mag of systematic uncertainty in our
distance moduli to account for the known source of uncertainty of extinction at the pole.

In most supernova work we assume the Galactic reddening law \citep{cardelli89}
applies to external galaxies ($R_V=3.1$), but studies of individual
\sneia\ have found a range of values extending to much smaller values of
$R_V$ \citep{riess96,tripp98,phillips99,krisciunas00,wang03,altavilla04,reindl05,elias-rosa06}.
These measurements are dominated by objects with large extinction
values, where a significant measurement can be made 
can be made of the extinction law (lessening the effects of intrinsic
color scatter and systematic color variations with luminosity), and it
is possible that $R_V$ is correlated with total extinction \citep{jha06c}.
In principle, with photometry in three or more passbands, it is possible
to fit for $R_V$, but in practice, at $z>0.2$, there are only a few 
\sneia\ in the literature with the requisite high-precision photometry extending
from the rest-frame UV to the near-IR. The systematic error on our
measurement of $D_L$ caused by assuming a particular value of $R_V$ depends on
the average extinction as a function redshift, assuming $R_V$ is
constant with $z$, except for a small correction caused by the rest-frame
effective bandpass of our filters drifting away from the low-$z$ values,
depending on the precise redshift of each object. To quantify this
effect, we fit our complete distance set with three different values of
$R_V$:  $2.1$, $3.1$, and $4.1$.

\subsection{Color and Extinction Distributions and Priors}
\label{sec:priors}

To evaluate the systematic effects produced by various prior
assumptions about extinction, we have fit the entire data set with a
variety of plausible priors: the ``exponential'' prior of \citet{jha06c},
a flat prior from $-\infty$ to $+\infty$ 
(the ``flatnegav'' prior), and an exponential prior with an
added Gaussian around zero that is based on models of the dust
distribution in galaxies (``glos'' and the redshift-dependent ``glosz'').
These results are presented in \S\ref{sec:cosmology} and form the basis for Table~\ref{tab:w_rv_sys}.

To separate the effects of color and extinction,
\citet{jha06c} noted that the distribution of color excess
in their nearby sample was consistent with a Gaussian distribution
of $\sigma=0.2$ convolved with a one-sided exponential, $\exp{(-A_V/\tau)}$,
where $\tau=0.46$~mag.
As discussed in \S\ref{sec:extinction_prior}, the ``glosz'' prior we
adopt here is derived from models of line-of-sight dust distributions
in galaxies.
It has more parameters than the simple exponential model of
\citet{jha06c}, but we believe these additional parameters are
well motivated.

The power of MLCS2k2 to distinguish between color and extinction lies in the
ability to treat the two phenomena independently.
A06 uses SALT and makes the assumption that the color$+$extinction distribution is the same in the nearby and in the high-redshift samples; 
the separation of the $A_V$ component in the MLCS2k2
model allows us to model our expected distribution of $A_V$
based on both models of dust in galaxies
and selection effects of the ESSENCE survey.
This separation allows us to take the nominal ``glos'' model and create
the ``glosz'' prior that combines the distribution of dust in galaxies
with the redshift-dependent selection effects.

The difference in the mean estimated parameter for a constant $w$ is
given in Table~\ref{tab:w_rv_sys} for the different MLCS2k2 $A_V$
priors discussed above.
For the main MLCS2k2 ``glosz''
analysis we present here, we find a slope of 
${\Delta}w / {\Delta}R_V=0.02$ in the dependence
of $w$ on the assumed $R_V$.
The effect on $w$ of
varying $R_V$ is substantially greater for the less restrictive
$A_V$ priors because the covariance between $A_V$ and $\mu$ is 
substantially greater for these priors.  
A reasonable variation of $0.5$ in the value for $R_V$
contributes a systematic uncertainty of ${\Delta}w=0.01$

Differences in the inferred value of $w$ for various assumed
absorption priors shows that this is a significant systematic effect.
The maximum
difference between two priors, ``exponential'' and ``glosz,''
for the nominal $R_V=3.1$ case is ${\Delta}w=0.165$.
While we have conducted careful simulations to determine the 
most appropriate prior for our sample (see \S\ref{sec:selection}) and it
is clear that the ``exponential'' is not appropriate for this analysis, 
we nonetheless take half of the difference between the two as
representative of our systematic uncertainty, 
$\Delta_w^{\rm prior}=0.08$, due to the choice of prior.
The residual $0.02$~mag shift of the simulations with the ``glosz'' prior
shown in Fig.~\ref{fig:cut} for $z\approx0.65$ results in a very small
shift in ${\Delta}w$ of only $0.001$.
Since we use an $A_V$, that obviously interacts strongly with our 
understanding of the intrinsic color distribution of \sneia.
We estimate this contribution to our systematic error budget at ${\Delta}w=0.06$
We have not undertaken a similar
analysis with the SALT fitter, but the underlying assumption
that the color, extinction, luminosity relationship for \sneia\ is constant
with redshift is subject to uncertainties analogous to those
considered here in the context of the MLCS2k2 $A_V$ prior.
The issue of color and extinction distributions clearly needs to be
addressed for substantial further progress to be made in the field of supernova cosmology.

\subsection{Malmquist Bias and Other Selection Effects}
\label{sec:malmquist}

As with all magnitude-limited surveys, at the faint limits of the
survey we are more likely to observe objects drawn from the bright
end of the \snia\ luminosity distribution. 
This Malmquist bias is particularly dangerous for inferences about cosmology based on supernova observations.  
However, it is not necessarily troubling that we may observe more
luminous, broad events at high redshift, as long as the known
empirical luminosity-width relation is valid at those redshifts. 
Rather, the concern for cosmological measurements is that at high
redshift, we may preferentially find \sneia\ which are bright \emph{for
their light curve shape}.
A second and more subtle concern is that at higher redshifts we are
also less likely to detect \sneia\ whose light suffers 
significant absorption due to dust in their host galaxies.

We have modeled both of these effects (see \S\ref{sec:selection} \& \ref{sec:extinction}) 
and have controlled for their impact.
Our current limits on systematics due to uncontrolled selection
effects is $\Delta_w^{\rm selection}=0.02$.  A thorough study of the
efficiency of the ESSENCE survey will be presented by 
\citet{pignata07}.  We aim for this future work to allow us to reduce this
contribution to our systematic error to no more than 1\%.

\subsection{Type Ia Supernova Evolution}
\label{sec:snevolution} 

A persistent concern for any standard-candle
cosmology is the possibility that the distant candles may differ slightly
from their low-redshift counterparts. 
In a recent paper \citep{blondin06}
we compare the spectra of the high-redshift \sneia\ in this sample
with low-redshift \sneia\ and demonstrate that there is no evidence
for any systematic difference in their properties. This conclusion is
based on line-profile morphology and measurements of the
phase-evolution of the velocity location of maximum absorption and
peak emission.

These results confirm a number of other studies of distant \sneia
\citep[e.g.,][]{coil00,sullivan03,lidman04} that all confirm that, to
the accuracy of current observations, the high and low redshift supernova
populations are indistinguishable. Recently \citet{hook05} used spectral
dating, spectral time sequences, and measurements of expansion velocities
to compare distant and nearby \sneia; they also find
no evidence for evolution in \snia\ properties up to $z\approx0.8$.

Although we are confident that the subtypes of distant \sneia\ are
well represented by the subtypes seen nearby, we cannot rule out a subtle
shift in the population demographics that may yet bias the estimates
of cosmological parameters. This potential bias is of particular concern for future
experiments that plan to measure the equation-of-state parameter, $w$, with an accuracy
of a few percent. There is now evidence that \snia\ properties are
correlated with host-galaxy morphology. 
\citet{hamuy96} and \citet{riess99} 
show that the brightest \sneia\ occur only in galaxies
with on-going star formation.
However, they observe no residual correlation after light-curve shape correction.
Because the galactic demographics over the redshift range of interest
change less than current variations in stellar population of \snia
host galaxies, we remain confident that our one-parameter correction
for supernova luminosity adequately corrects any shift in the average
luminosity of \sneia\ to the same precision as in the nearby Universe,
$\sigma_\mu < 0.02$~mag.   
We thus estimate a systematic uncertainty from possible \snia
evolution on our measurement of $w$ of ${\Delta}w=0.02$.

One way to verify this confidence is to search for additional parameters that
allow tighter luminosity groupings of the low-redshift population.
In a first, reassuring step, Hubble diagrams for subsets of
\sneia\ based on host-galaxy type separately confirm the accelerated
expansion of the Universe~\citep{sullivan03}.

\subsection{Hubble Bubble and Local Large-Scale Structure}
\label{sec:bubble}

The local large-scale structure and associated correlated flows of the
Universe should not yet present a significant contribution to the
systematic error budget of the current survey~\citep{hui06,cooray06}.
However, at the lowest multipoles we are sensitive to local correlated
flows, and, at the most extreme, our cosmological results would be 
sensitive to a local velocity monopole or ``Hubble bubble.''  
\citet{jha06c} see such an effect in their analysis of nearby \sneia.
We use only the subset of \sneia\ from \citet{jha06c} with $z>0.015$
and find that this effect could contribute as much as $0.065$ to our
systematic error budget in $w$.
We will rely on future sets of nearby \sneia\ ($0.01<z<0.05$) that are now being
acquired at the CfA, by the Carnegie Supernova program, 
by the Lick Observatory Supernova Search,
and by the
SNfactory to reduce this uncertainty below 2\% to help achieve the
desired systematic uncertainty required for the final ESSENCE analysis.

\subsection{Gravitational Lensing}
\label{sec:lensing}

Gravitational lensing can increase or decrease the observed flux from
a distant object.  
The expected distribution is asymmetric about the average 
flux multiplier of unity.
\citet{holz05} calculate the effect for \snia
surveys and determine that any systematic effect from neglecting the
asymmetry of the probability distribution function for magnification (as we do here) decreases quickly
with the number of \sneia\ per effective bin.  Roughly speaking, at a
$z\approx0.5$,  in a redshift bin width of
${\Delta}z \sim 0.1$, ten \sneia\ per bin are sufficient to reduce any systematic effect
in luminosity distance to less than $0.3\%$, which makes no noticeable contribution to 
our systematic error budget.  
For the redshifts of interest in the ESSENCE survey, lensing has a
more significant effect in the scatter it adds to the observed
brightness of \sneia.
\citet{holz05} calculate a 3\% increase in the dispersion in distance
modulus at $z\approx0.5$.
We include the effect of lensing in our analysis by adding a statistical dispersion of 
$\sigma^{\rm lensing}_{\mu}=0.03$ to our luminosity distance uncertainty
for the ESSENCE and SNLS \sneia.

\subsection{Grey Dust}
\label{sec:greydust}

When the first cosmological results with \sneia\ were announced,
that distant \sneia\ were dimmer than they would be in a
decelerating Universe, \citet{aguirre99a,aguirre99b} suggested various models
for intergalactic grey dust that could explain this dimming without producing observable reddening.  
To explain \sneia
becoming consistently dimmer with distance, this dust would need to be
distributed throughout intergalactic space beginning at
least at $z=2$~\citep{goobar02}.  The most
naive model of such dust distribution and creation
would predict that \sneia\ should continue to get dimmer
relative to a flat, \OM$=1$, cosmology all the way up to
at least a redshift of $2$.  The high-redshift \snia\ work of
\citet{riess04b} demonstrated that this continued dimming is not what is observed:
the apparent magnitudes of \sneia\ become first a little dimmer and then
a little brighter with redshift than they would in an empty Universe. 
This is exactly what we expect from an early phase of deceleration
followed by a recent phase of acceleration in a mixed, dark-matter/dark-energy cosmology.

A more complicated model of dust was contrived by \citet{goobar02}.
It involves the creation of intergalactic dust at just the right rate
to match the decrease in opacity due to expansion of the Universe. 
This carefully constructed
model mimics the signal of an accelerating universe and is difficult
to distinguish from a universe that is presently dominated by dark
energy.  This model does not have a strong underpinning in the
behavior of known dust and represents a form of fine-tuning.  In the
larger context of converging cosmological evidence, this particular
scheme for matching the data seems less plausible than a universe with
dark energy.

Recent observational constraints from non-\snia\ sources have
independently placed significant constraints on the amount of
intergalactic dust~\citep{petric06,ostman06}.
In particular, the observations of \citet{petric06} limit
intergalactic dust to contributing no more than one percent 
to potential dimming of light out to a redshift of $0.5$, based on upper limits
to X-ray scattering by dust along the line of sight to a quasar at $z=4.3$.

\pagebreak
\pagebreak
\section{Cosmological Results from the ESSENCE Four-Year Data}
\label{sec:cosmology}

The ESSENCE \sneia\ allow us test the hypothesis of a \lcdm\ concordance
model and constrain flat, constant-$w$ models of the Universe.
We use our MLCS2k2 light-curve fitting technique to measure luminosity
distances to nearby and ESSENCE \sneia\ (Table~\ref{tab:mlcs_fits_prior_glosz}).
When then fit cosmological models to constrain the properties
of the dark energy.
We compare the results we obtain using MLCS2k2 with 
those obtained using the SALT light-curve fitter~\citep{guy05}.
The SALT fitter was used to fit the nearby light curves, our ESSENCE
light curves, and the SNLS
light curves.\footnote{\url{http://snls.in2p3.fr/conf/release/}}
To verify that we were making appropriate use of the fitter, we
fit the nearby and SNLS light curves with SALT, taking the same
$\alpha=1.52$ and $\beta=1.57$ width and color parameters used in
A06.  We recovered the $w$ result of
A06 to within $0.01$ in best-fit constant $w$ in a model
with a flat Universe
using the cosmology fitter that we employ here\footnote{\url{http://qold.astro.utoronto.ca/conley/simple\_cosfitter/}}.
We have compiled our light curves of nearby \sneia\ from the literature
independently of the SNLS analysis and used slightly different
quality cuts, so it is quite
encouraging that we can replicate these results.
Table~\ref{tab:salt_fits} gives the SALT fit parameters for the
nearby, ESSENCE, and SNLS \sneia\ discussed here.

\subsection{ESSENCE \snia\ Sample}
\label{sec:essence}

For the ESSENCE project we find that using 
photometric selection criteria based on the color and
rise time of the candidate object, similar to those used by the
SNLS~\citep{howell05,sullivan06a}, and in good weather and seeing conditions, 80\% of the candidates we take spectra of are \sneia.
We use a deterministic analysis~\citep{blondin07}, as described in
\citet{miknaitis07}, to determine final types and redshifts for our
SNe and to cull objects that are not \sneia\ from our sample.
All of the ESSENCE supernovae used in this analysis were spectroscopically
confirmed as \sneia{}.

From 2002--2005 the ESSENCE project discovered and spectroscopically
confirmed 113~\sneia{}.  As discussed by \citet{miknaitis07}, 
which gives full details of these \sneia\ including their RA and Dec, only
4 of the 15 \sneia\ from 2002 have been fully analyzed so that leaves
us with 102~\sneia.
Although we kept 91T-like \sneia\ such as d083,
d085, and d093, we rejected the peculiar \snia{}
d100~\citep{matheson05}.  Three \sneia\ were rejected from the
nearby+ESSENCE only fits because they were at redshifts greater than
$0.67$ (see below).  After we applied the cuts in
Tables~\ref{tab:mlcs_cuts} and Tables~\ref{tab:salt_cuts}, we were left
with 57 and 60~\sneia\ for MLCS2k2 and SALT respectively.

With the
MLCS2k2 fitter, the largest cut was the 32 \sneia\ rejected because
they had fewer than 8 data points with an SNR $> 5$, 
no such points after +9 days, or no such points before +4 days.
Two of the 102~\sneia\ were
located near edges of the detector field-of-view that we later
determined were photometrically less reliable.  Due to high \chisqnu{}
or related poor light-curve goodness-of-fit values, 
we eliminated an additional 6~\sneia.  This left us with
a total of 57~\sneia\ for our main MLCS2k2 nearby+ESSENCE analysis.
The SALT fitter successfully fit three more \sneia\ than MLCS2k2, but,
in general, our SALT quality cuts accepted the same \sneia\ as our
MLCS2k2 quality cuts.  
The requirements we imposed here on the light curves were stringent
cuts to ensure reliable fit parameters.
We are currently engaged in an active program to improve the
sensitivity of \snia\ light-curve fitters and we anticipate recovering
50\% of the \sneia\ rejected here in the final ESSENCE analysis.

\subsection{Nearby \snia\ Sample}
\label{sec:nearby}

The \snia\ cosmological measurement is fundamentally a comparison of the
luminosity distance vs. redshift relation at low redshift and high
redshift.
The ESSENCE \sneia\ alone provide a homogeneous set of
luminosity distance vs. redshift measurements covering the redshift range 
$0.15<z<0.7$.
We selected our nearby \sneia\ from the set compiled by \citet{jha06c}.
We applied the light-curve criteria from 
Tables~\ref{tab:mlcs_cuts} and \ref{tab:salt_cuts} 
and also rejected known peculiar \sneia\ such as
SN~2000cx~\citep{li01} and SN~2002cx~\citep{li03,jha06b}.  
Our list of nearby \sneia\ has 41 \sneia\ in common
with the set used by A06.
To minimize complications from loosely constrained
local peculiar and coordinated flows, we only considered \sneia{}
with CMB-frame redshifts of $z>0.015$. Our final sample consists of 
\numnearbysne~nearby \sneia\ as listed in the fit parameter tables 
(Tables~\ref{tab:mlcs_fits_prior_glosz} and \ref{tab:salt_fits}).
We used the re-derived Landolt/Vega calibration of A06 to determine
the passbands for this set of nearby \sneia.
The light curves we used for these \sneia\ are also included with
the ESSENCE light curves available on our website.\footnote{\url{http://www.ctio.noao.edu/essence/}}

\subsection{External Constraints}

To provide complementary cosmological constraints on our \snia\ observations,
we include external information from baryon acoustic oscillations~\citep[BAO;][]{eisenstein05}.
The BAO constraints on (\OM, $w$) from \citet{eisenstein05} are the
most complementary measurement in the (\OM, $w$) plane to our \snia{}
measurements, relying only on the observed redshift and angular size
of the first doppler peak in the CMB and not on $H_0$. In addition,
because the BAO constraints on \OM\ are similar in precision (and
value) to those derived from large scale structure
\citep{percival01,percival02}, WMAP directly \citep{spergel06}, and
from the study of X-ray clusters \citep[for a review see][]{voit05},
we choose to combine our results only with the BAO results.

We compare the
specific model of a flat Universe with either $w=-1$ or constant $w$ of any value
to our data.  
\sneia\ have very little leverage on the global flatness of the Universe
because they effectively measure the difference between \OM\ and \OL,
and flatness depends on the sum.
\citet{eisenstein05} have constrained curvature to be within
\OK$=\pm0.01$ of flat. The results presented here (from the \sneia) on
$w$ are not significantly affected by variation of \OK\ by this amount,
because the effects of curvature are not noticeable until looking back
to much higher redshift. However, non-flat models will significant
alter the BAO results on (\OM, $w$) and therefore our joint
constraints.

For our analysis of the ESSENCE and nearby \sneia, 
we have chosen to additionally
limit our redshift range to $z<0.670$ to avoid using the rest-frame $U$ band.  Since this remove just three ESSENCE \sneia\ from our sample, the tradeoff is worthwhile
to minimize this source of uncertainty (see \S\ref{sec:distances}).  
When we add in the SNLS or Riess gold samples, we relax this
constraint to incorporate those higher-redshift \sneia.

In Figs.~\ref{fig:joint_mlcs_prior_glosz_hubble_diagram} and
\ref{fig:joint_salt_hubble_diagram} we show Hubble diagrams of the nearby,
ESSENCE, and SNLS samples for the two different fitters we consider in this
paper.  We overplot an empty Universe (\OM,\OL,$w$) = $(0,0,-1)$, a
matter-only open Universe $(0.3,0,-1)$, and a \lcdm\ concordance cosmology
$(0.27,0.73,-1)$.  The residuals in luminosity distance are then
shown with respect to the \lcdm\ model.  MLCS2k2 appears to be more
suited for the ESSENCE data sample than SALT, although the latter
benefits from its flux-based fitting by being able to extract useful
luminosity distances from a few more \sneia.  One \snia, ``d083,'' is a
particular outlier in both fitters at $\sim0.5$~mag brighter than
expected in the best-fit or \lcdm\ cosmologies.  
\citet{matheson05} found the spectrum of this object to be like that
of SN~1991T, which is the archetype of over-luminous \sneia.
This \snia\ is likely an
interesting object worthy of further study and is potentially
similar to a similarly super-luminous object, SN~2003fg, found in the SNLS survey~\citep{howell06}.  However, given that our
sample comprises \numWsne\ objects, we certainly allow for the
reasonable statistical possibility of a 3$\sigma$ outlier such as ``d083''
and thus retain it in our sample.

In Fig.~\ref{fig:mlcs_prior_glosz_salt_om_w} we show
the 1$\sigma$, 2$\sigma$, and 3$\sigma$ probability contours for our measurement of $w$ vs. \OM\ for
ESSENCE+nearby alone, the BAO constraints from \citet{eisenstein05},
and the combination of the \snia\ and BAO constraints.

Table~\ref{tab:results} shows the cosmological parameters $w$ and
\OM\ for each of these sets for flat models of the Universe with a constant $w$ 
as well as the \chisqnu\ for a concordance cosmology and the 1-D marginalized values.
A \lcdm\ model of the Universe fits the
MLCS2k2-analyzed ESSENCE+nearby sample with a \chisqnu\ of \chisqnugloszlcdm
and a residual standard deviation of \stddevgloszlcdm~mag.
Thus, while the estimated $w$ parameter in the constant-$w$ models is
$w=$\wglosz, a flat, $w=-1$ model of the Universe is consistent with our
data.

Our results from these \numWsne~\sneia\ from the ESSENCE survey are
consistent with the results of A06.
It is reassuring that two independent teams using different telescopes
and studying different regions of the sky find that \sneia\ at high
redshift exhibit the same luminosity distance vs. redshift relationship.
These samples strengthen and extend the evidence from \sneia\ for 
dark energy and, together with complementary constraints on \OM,
point toward simple \lcdm\ models for our Universe.

\subsection{Joint ESSENCE+SNLS Cosmological Constraints}

A new opportunity presents itself with the release of the SNLS
light curves from A06 and the light curves presented in
this paper.  For the first time it is possible to do a proper, self-consistent
joint fit of two large, independent sets of distant \sneia.

When fitting the SNLS \sneia\ with MLCS2k2 and the ``glosz'' prior we
shift the assumed $A_V$ and $\Delta$ prior selection window functions
by ${\Delta}z=+0.20$ to represent the greater depth of the SNLS survey.
The proper way to derive this prior for SNLS would be to
model the SNLS survey efficiency and and fit simulated \sneia\ with
MLCS2k2 as we presented in \S\ref{sec:selection} for the ESSENCE
survey.
Similar concerns apply for possible selection effects in the
heterogeneously nearby sample.
Nevertheless, we believe our use of the ``glosz'' prior is appropriate
for the low-redshift sample (where it is just the ``glos'' prior) and
the simple extension in redshift to be a reasonable approach for the
SNLS sample.
The additional systematic errors introduced by this joint comparison
center on the photometric calibration of the distant sample relative
to the nearby \sneia.  We estimate that uncertainty to be the same as
the calibration uncertainty to the nominal Vega system used by each
project: $\Delta{\rm zpt}=0.02$~mag.  We have not modeled different
offsets between the two data sets, but merely express the uncertainty
as an additional uncertainty in our inferred cosmological
parameters.
This relative zeropoint uncertainty adds an additional ${\Delta}w=0.02$
to our overall systematic uncertainty on $w$.

With our combined analysis, we start with the traditional \OM-\OL
contour plot that was the first clear evidence for dark energy.

Table~\ref{tab:joint_results} shows the cosmological parameters $w_0$ and
\OM\ for each of these sets for flat models of the Universe with a constant $w$
as well as the \chisqnu\ for a concordance cosmology.
A \lcdm\ model of the Universe fits the
SNLS+ESSENCE+nearby sample analyzed using MLCS2k2 ``glosz'' with a 
\chisqnu\ of \chisqnugloszjointlcdm\ from \glosznumsnejoint~\sneia
and a residual standard deviation of \stddevgloszjointlcdm~mag.  
A joint analysis of the luminosity distances from the SALT fitter results in a 
\chisqnu\ of \chisqnusaltjointlcdm\ from \saltnumsnejoint~\sneia
and a residual standard deviation of \stddevsaltjointlcdm~mag..
Fig.~\ref{fig:joint_mlcs_prior_glosz_salt_om_w} 
show the joint MLCS2k2 and SALT results for this joint sample.
The estimated $w$ parameter in the constant-$w$ models is
$w=$\wgloszjoint, and a flat, $w=-1$ model of the Universe 
remains consistent with the current generation of \snia\ data.

\subsection{Joint ESSENCE+SNLS+Riess Gold Sample Cosmological Constraints}

In order to explore models with varying $w$, we now include the gold
sample from \citet{riess04b} to extend our reach out to $z\approx1.5$.
The high-quality intermediate-redshift samples of the ESSENCE and SNLS
surveys provide an excellent complement to the high-redshift \sneia
in this set.
The heterogeneous nature of the collection of \sneia\ in the
gold sample makes it beyond the scope of this paper to produce 
definite estimates of the systematic errors that result from including
this additional set, but
it is tempting to add these \sneia\ and examine the new
constraints on cosmological parameters.

We used the 39 nearby \sneia\ in common between the nearby \snia\ sample
we discuss here and the gold sample to normalize the
luminosity distances between the two sets.
To avoid double-counting of \sneia\ in this joint analysis, we then
drop the nearby \sneia\ from the gold sample and use only the nearby
\sneia\ fit in this paper.

We first compute the \OM-\OL\ contours to update the case for dark energy
from \sneia.
Fig.~\ref{fig:OMOL_OMw_joint_riess04} represents
the most stringent demonstration to date of the existence of dark energy.
The \sneia\ data alone rule out an empty Universe at
$4.5$~$\sigma$, an (\OM, \OL) = $(0.3, 0)$ Universe at $10$~$\sigma$,
and an (\OM, \OL) = $(1, 0)$~$\sigma$ Universe at $>20$~$\sigma$.
The joint constraints on constant-$w$ models from this full set are
$w=-1.09^{+0.09}_{-0.10}$.  The highest-redshift data do not noticeably
improve constraints for these models over the set of intermediate-redshift \sneia\ from ESSENCE+SNLS.  
It is for models with variable $w$ that the high-redshift data summarized by \citet{riess04b} provide the most utility.
We here provide the global constraints on models characterized by
$w=w_0+w_a(1-a)$~\citep{linder03,albrecht06}.  
Using the BAO constraints on variable $w$ models would require
integration from $z=0.35$ to $z\sim1089$ and the corresponding
assumption that $w=w_0+w_a(1-a)$ is the proper parameterization over
this stretch.  If one is willing to make this assumption, then BAO+CMB
already places significant constraints on the allowed $(w_0,w_a)$
parameter space.  However, given that our multi-variable
parameterizations of $w$ are arbitrary models with no clear
theoretical motivation, we instead choose to regard $w=w_0+w_a(1-a)$
as a local expansion valid out to a redshift of $\sim2$ but not
necessarily to $z\sim1089$.
We then explicitly assume \OM$=0.27\pm0.03$.
Fig.~\ref{fig:w0wa_joint_riess04} shows the $(w_0, w_a)$ contours for
this combined analysis.  
These constraints represent the advances of our understanding of dark
energy.  It is clear that work remains to constrain models of
variable $w$.

\pagebreak

\section{Conclusions}
\label{sec:conclusions}

The ESSENCE survey has successfully discovered, confirmed, and
followed 119~\sneia\ in our first four years of operation.  We
presented results from an analysis of \numWsne\ of those \sneia\ here, chosen
so as to maximize insight while minimizing susceptibility to
systematic errors. We have expended considerable effort to make
quantitative estimates of various sources of systematic uncertainty
that may afflict the ESSENCE results; of these, host-galaxy
extinction and a potential local velocity monopole 
are currently the predominant concerns. We are working to
devise ways to better estimate extinction, using both spectroscopic
and photometric observations.  Ideally, we would use all available information 
to arrive at an extinction prior customized for each supernova 
(e.g., different priors for elliptical and spiral host galaxies), 
rather than applying a single prior to the collection of all light
curves.

The ESSENCE photometric calibration uncertainties will be reduced by
an intensive calibration campaign this fall on the CTIO 4-m telescope in
conjunction with the improved calibration of the SDSS southern stripe
from the SDSS~II project~\citep{frieman04,dilday05}.  We hope to reduce our overall
systematic uncertainty to the 5\% level through this improved
photometric calibration and an improved external nearby
\snia\ sample from KAIT, the Nearby Supernova Factory, CfA, SDSS~II, and the Carnegie SN Program
to reduce our systematic sensitivity to a potential velocity monopole
in the local \snia\ sample.

Combining our \snia\ observations with the BAO results of
\citet{eisenstein05} we find that a fit to a constant-$w$, flat-Universe 
model to our data finds an estimated parameter value of $w=$\wglosz\ with
a \chisqnu$=$\chisqnuglosz\ using our full set analyzed with the
MLCS2k2 fitter of \citet{jha06c}.  A $w=-1$, flat-Universe model is
consistent with our data.  
A combined
analysis of ESSENCE+SNLS+nearby results in a estimated mean parameter
of $w=$\wgloszjoint.
We have no reliable estimate of the systematic effects involving the SALT fitter 
but take our general systematic uncertainty of \wgloszsys\ as representative
of the issues currently confronting supernova cosmology.

The statistical increase from the \sneia\ from the entire 6-year ESSENCE
data set plus improved photometric calibration of our detector and
photometric measurements will reduce our statistical uncertainty to
7\% and, together with an improvement in our systematic uncertainties to the level 5\%,
allow us to surpass our goal of a 10\% measurement of a
constant $w$ in a flat Universe.

Establishing the nature of dark energy is among the most pressing
issues in the physical sciences today. The emerging impression that
the equation-of-state parameter is close to $w=-1$ makes it difficult to
determine whether the underlying physics arises in the particle
physics sector or from the classical cosmological constant of general
relativity. A value of $w=-1$ is perhaps the least informative
possible outcome. 
In our view, this state of affairs motivates a vigorous effort to push
the observational constraints to improve our sensitivity to the value
and derivative of $w$ and strongly encourages searching for other
indications of new physics, as we well may need another piece 
to solve the puzzle handed us by Nature.

\section{Acknowledgments}

Based in part on observations obtained at the Cerro Tololo
Inter-American Observatory, which is operated by the Association of
Universities for Research in Astronomy, Inc. (AURA) under cooperative
agreement with the National Science Foundation (NSF); the European
Southern Observatory, Chile (ESO Programmes 170.A-0519 and 176.A-0319); the Gemini
Observatory, which is operated by the Association of Universities for
Research in Astronomy, Inc., under a cooperative agreement with the
NSF on behalf of the Gemini partnership: the NSF (United States), the
Particle Physics and Astronomy Research Council (United Kingdom), the
National Research Council (Canada), CONICYT (Chile), the Australian
Research Council (Australia), CNPq (Brazil), and CONICET (Argentina)
(Programs  GN-2002B-Q-14, GS-2003B-Q-11, GN-2003B-Q-14, 
GS-2004B-Q-4, GN-2004B-Q-6, GS-2005B-Q-31, GN-2005B-Q-35); 
the Magellan
Telescopes at Las Campanas Observatory; the MMT Observatory, a joint
facility of the Smithsonian Institution and the University of Arizona;
and the F. L. Whipple Observatory, which is operated by the
Smithsonian Astrophysical Observatory. Some of the data presented
herein were obtained at the W. M. Keck Observatory, which is operated
as a scientific partnership among the California Institute of
Technology, the University of California, and the National Aeronautics
and Space Administration; the Observatory was made possible by the
generous financial support of the W. M. Keck Foundation.

The ESSENCE survey team is very grateful to the scientific and
technical staff at the observatories we have been privileged to use.

{\it Facilities:} 
  \facility{Blanco (MOSAIC II)}, 
  \facility{CTIO:0.9m (CFCCD)}, 
  \facility{Gemini:South (GMOS)}, 
  \facility{Gemini:North (GMOS)}, 
  \facility{Keck:I (LRIS)},
  \facility{Keck:II (DEIMOS, ESI)},
  \facility{VLT (FORS1)},
  \facility{Magellan:Baade (IMACS)}, 
  \facility{Magellan:Clay (LDSS2)}.

The survey is supported by the US National Science Foundation through
grants AST-0443378, AST-057475, and AST-0607485.
The Dark Cosmology Centre is funded by the Danish National Research
Foundation.
SJ thanks the Stanford Linear Accelerator Center for support via a Panofsky Fellowship.
AR thanks the NOAO Goldberg fellowship program for its support.
PMG is supported in part by NASA Long-Term Astrophysics Grant NAG5-9364 and NASA/HST Grant GO-09860.
RPK enjoy support from AST06-06772 and PHY99-07949 to the Kavli Institute for Theoretical Physics.
AC acknoledges the support of CONICYT, Chile, under grants FONDECYT 1051061 and FONDAP Center for Astrophysics 15010003.

Our project was made possible by the survey program administered by
NOAO, and builds upon the data reduction pipeline developed by the
SuperMacho collaboration.
IRAF is distributed by the National Optical Astronomy Observatory,
which is operated by AURA under cooperative agreement with the NSF.
The data analysis in this paper has made extensive use of the Hydra
computer cluster run by the Computation Facility 
at the Harvard-Smithsonian Center for Astrophysics.
We also acknowledge the support of Harvard University. 

This paper is dedicated to the memory of our friend and colleague Bob Schommer.

\bibliographystyle{apj}
\bibliography{apj-jour,cos_paper}

\pagebreak

\begin{figure}
\plottwo{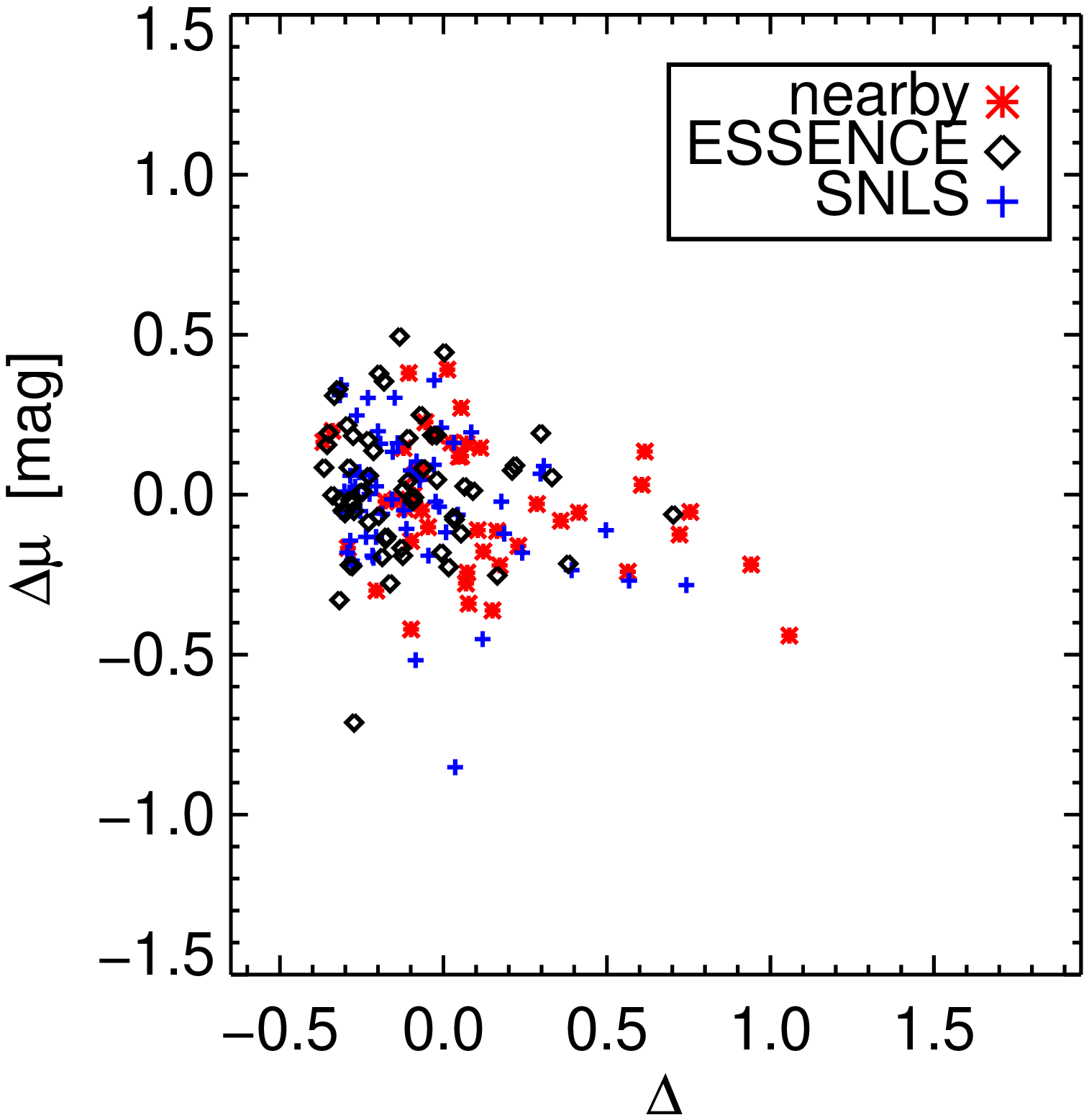}{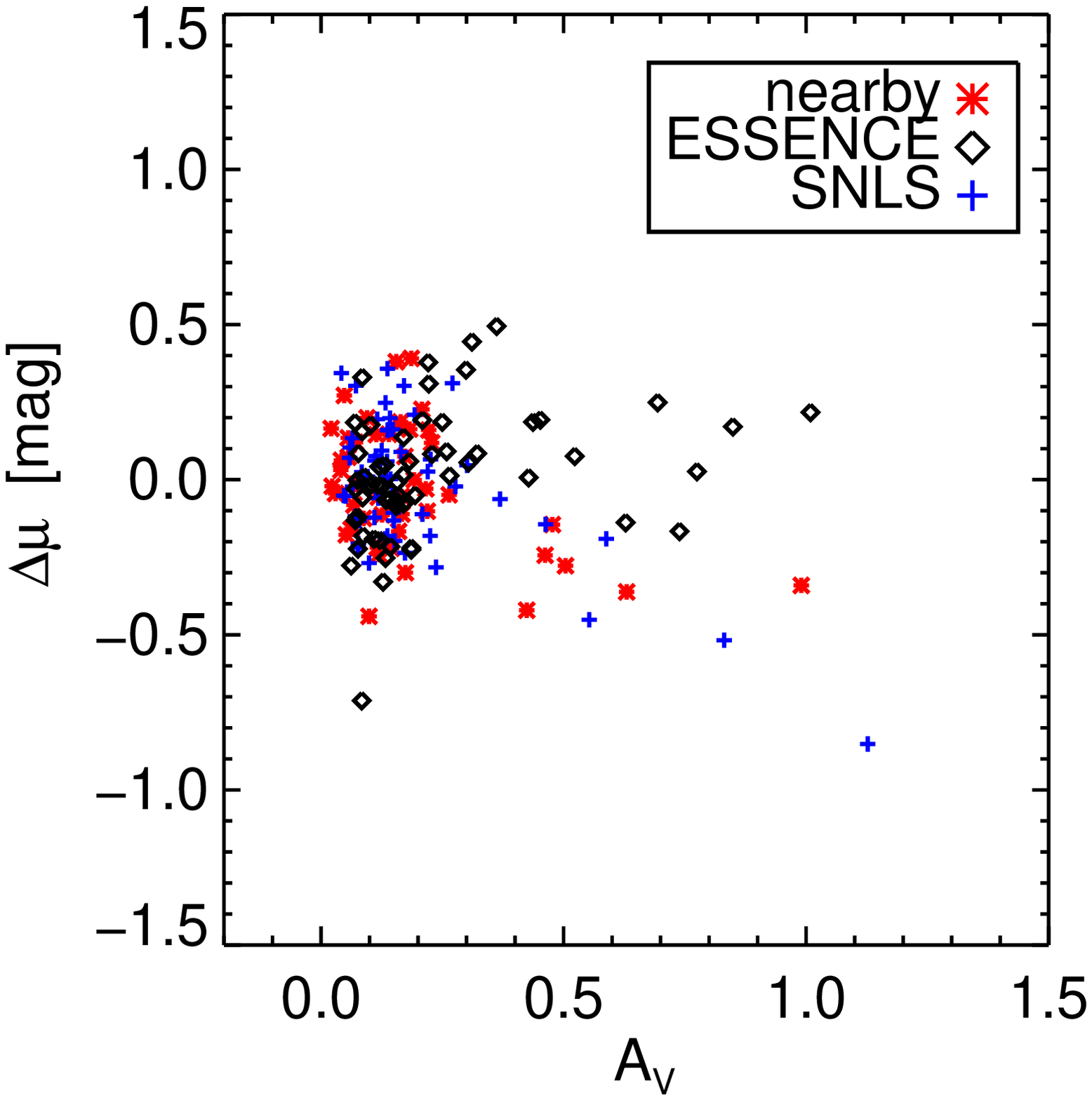}
\caption{
Distance modulus, $\mu$, residuals with respect to a \lcdm\ cosmology as a function of the MLCS2k2
``glosz'' fit parameters: $\Delta$ and $A_V$.  
See Table~\ref{tab:mlcs_fits_prior_glosz}.
}
\label{fig:mlcs_fits_resid_prior_glosz}
\end{figure}

\begin{figure}
\plottwo{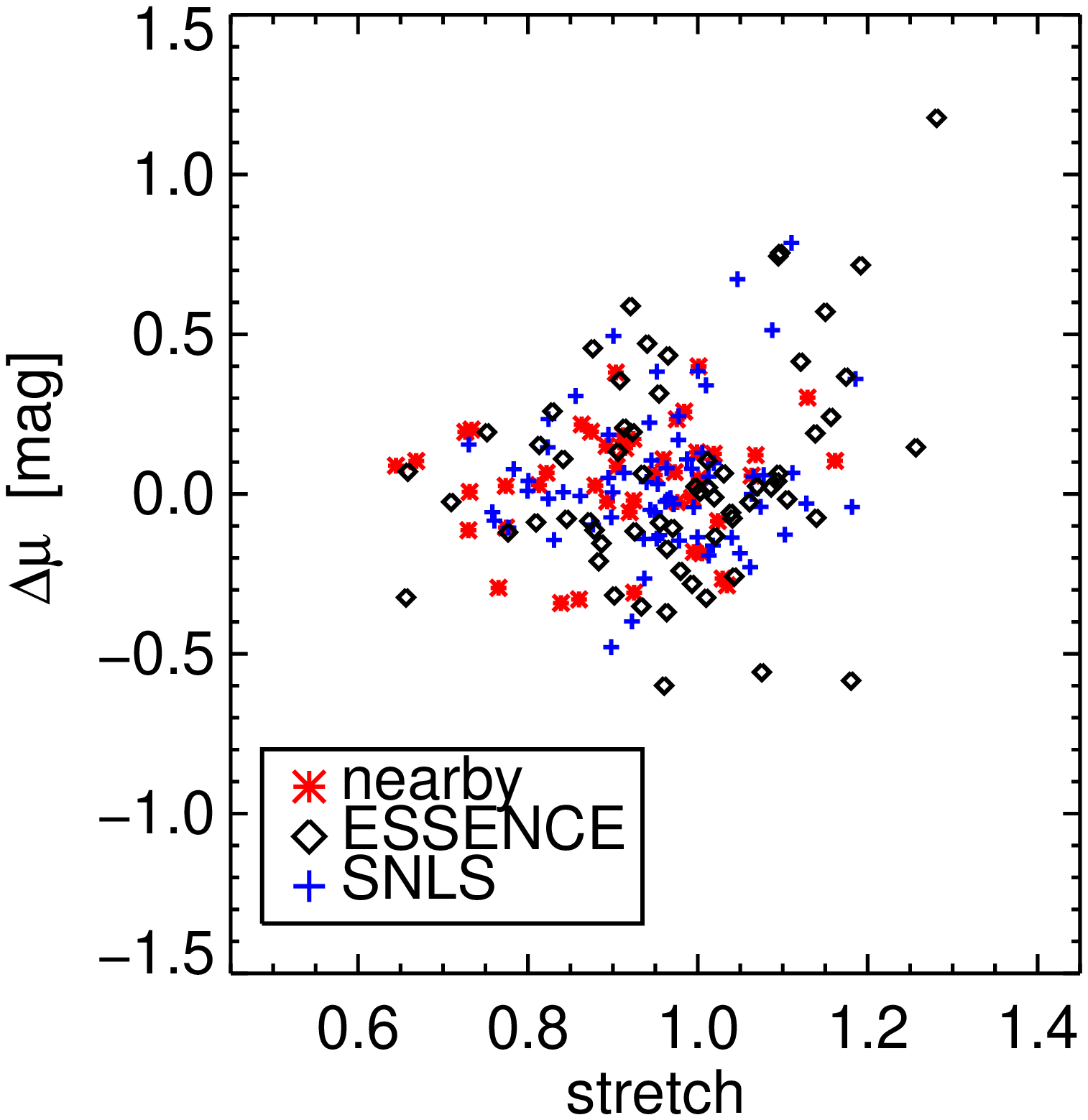}{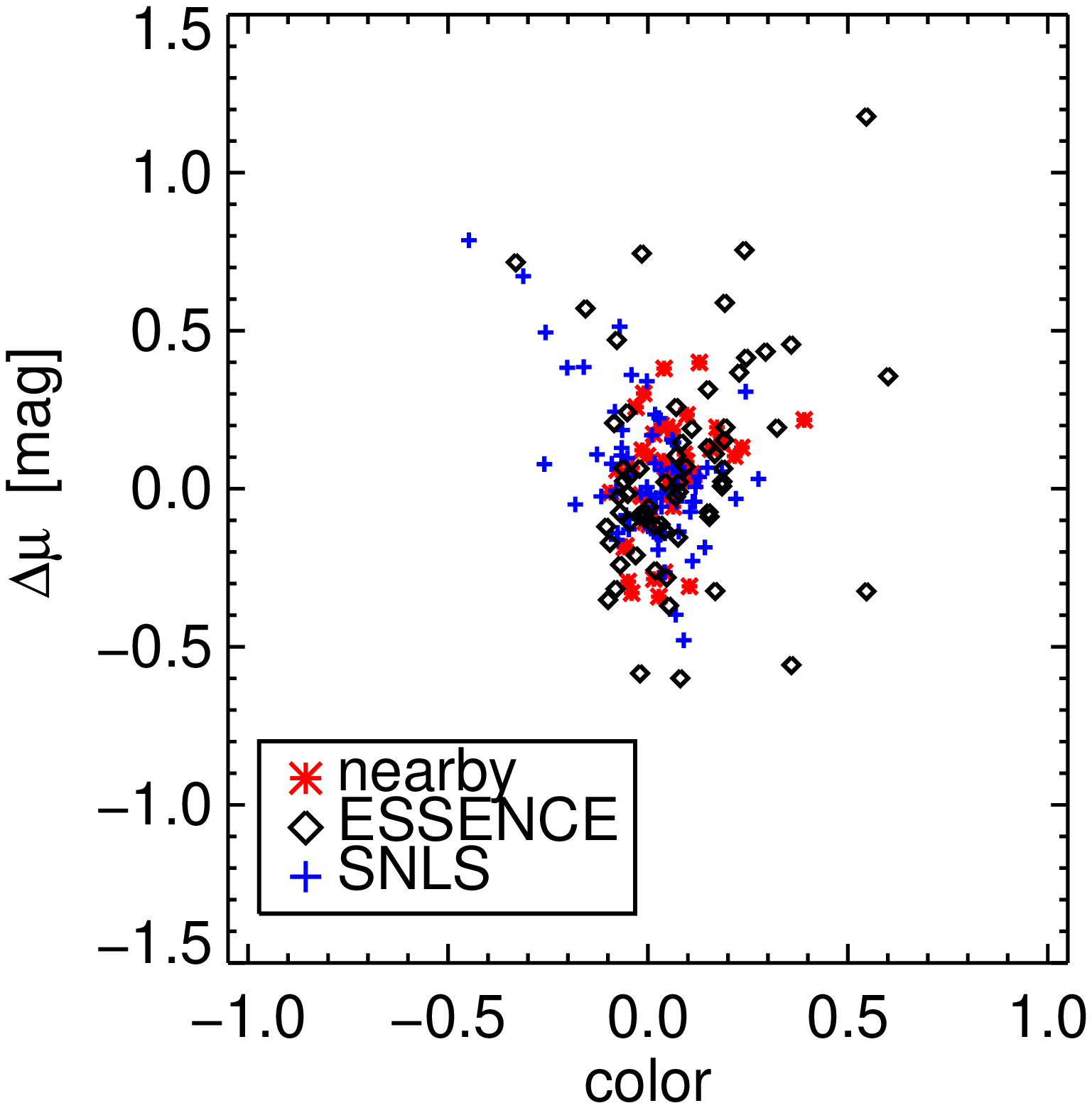}
\caption{
Distance modulus, $\mu$, residuals with respect to a \lcdm\ cosmology as a function of the SALT fit
parameters: stretch and color.  
See Table~\ref{tab:salt_fits}.
}
\label{fig:salt_fits_resid}
\end{figure}

\begin{figure}
\plotone{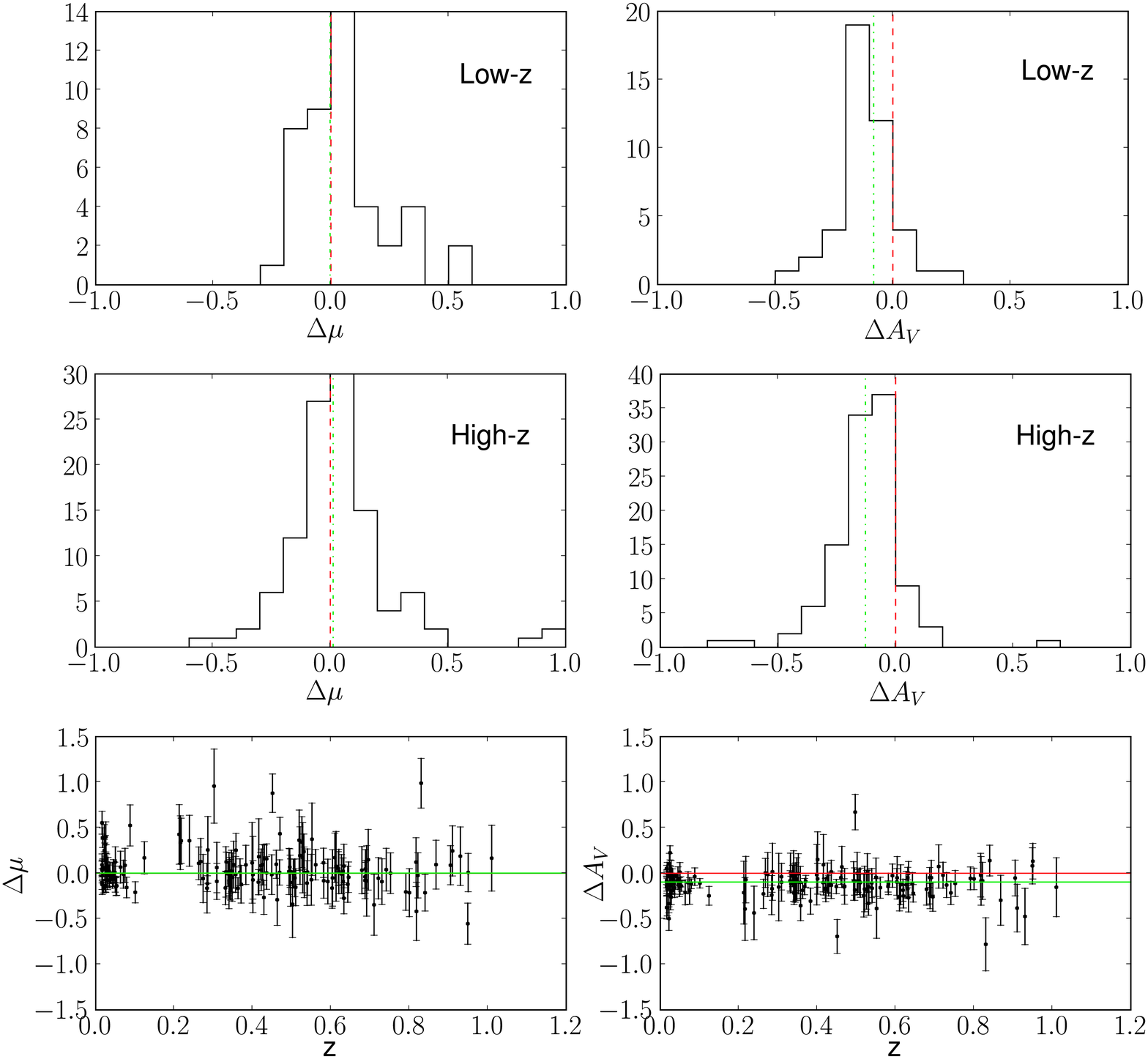}
\caption{
The distance modulus and $A_V$ as a function of redshift 
for MLCS2k2 ``glosz'' minus the SALT distance modulus and $\beta\times{\rm color}$ ($\beta=1.57$) for the
ESSENCE, SNLS, and nearby data sets.  
High-z refers to \sneia\ with $z\ge0.15$, low-z to $z<0.15$.
The dot-dash line shows the weighted average of the difference for each quantity
while the dashed line shows the line of zero difference.  
While the luminosity distances are offset between the two fitters, this
is mainly due to a slightly different definition of the \scriptM
parameter that defines the absolute luminosity of a \snia\ and the
Hubble constant.
The relative average difference between low redshift and high redshift
is $-0.0023$ mag.
This agreement translates to a similar agreement in the cosmological
parameters obtained with each approach 
(see Figs.~\ref{fig:mlcs_prior_glosz_salt_om_w} and 
\ref{fig:joint_mlcs_prior_glosz_salt_om_w}).
}
\label{fig:mlcs_vs_salt}
\end{figure}

\pagebreak
\clearpage

\begin{figure}
\plotone{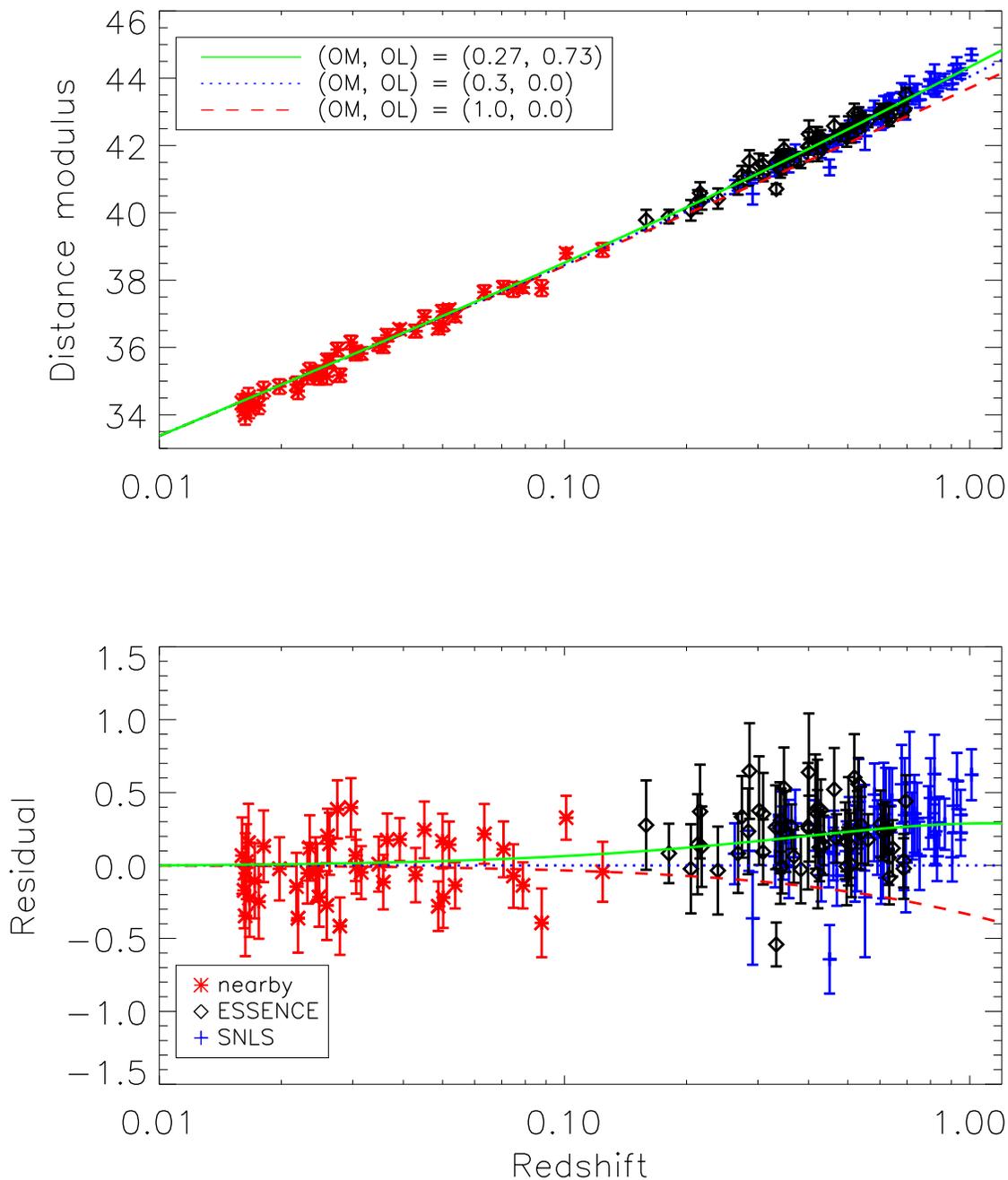}
\caption{Luminosity distance modulus vs. redshift for the ESSENCE, SNLS, and nearby \sneia\ for MLCS2k2 with the ``glosz'' $A_V$ prior.  For comparison the overplotted solid line and residuals are for a \lcdm\ ($w$, \OM, \OL) = $(-1, 0.27, 0.73)$ Universe.}
\label{fig:joint_mlcs_prior_glosz_hubble_diagram}
\end{figure}

\begin{figure}
\plotone{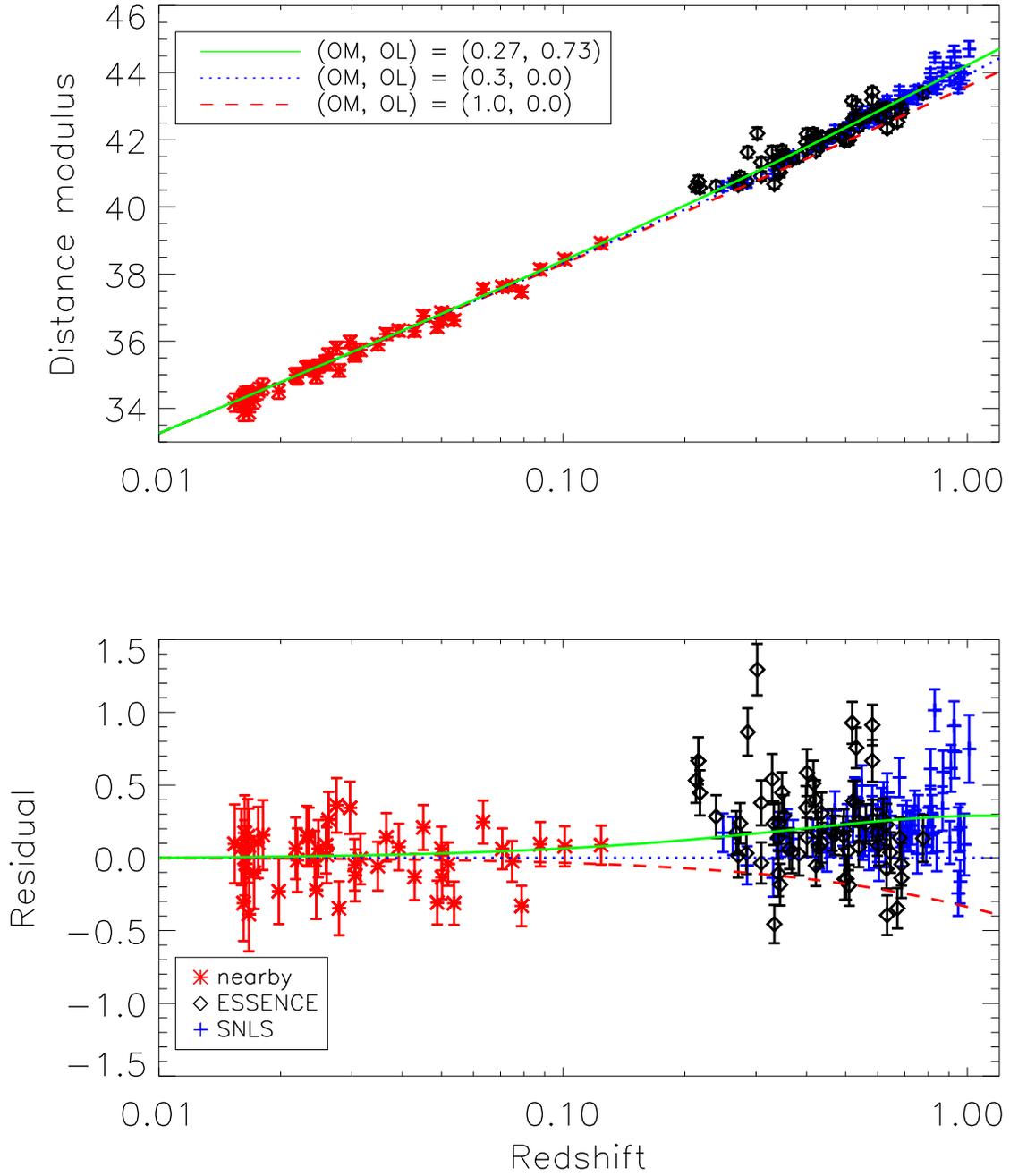}
\caption{Luminosity distance modulus vs. redshift for the ESSENCE, SNLS, and nearby \sneia\ for SALT.  For comparison the overplotted solid line and residuals are for a \lcdm\ ($w$, \OM, \OL) = $(-1, 0.27, 0.73)$ Universe.}
\label{fig:joint_salt_hubble_diagram}
\end{figure}

\begin{figure}
\plottwo{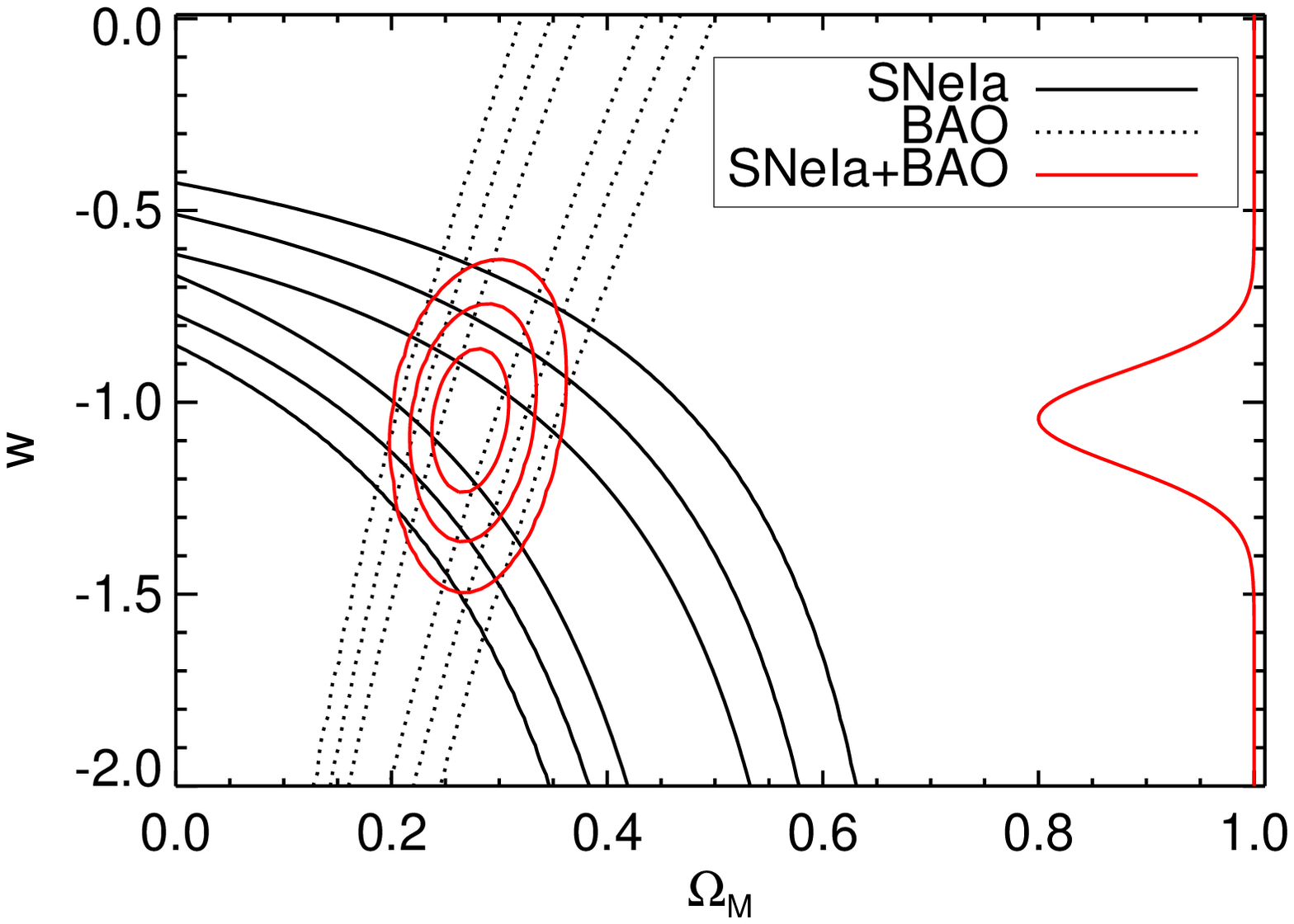}{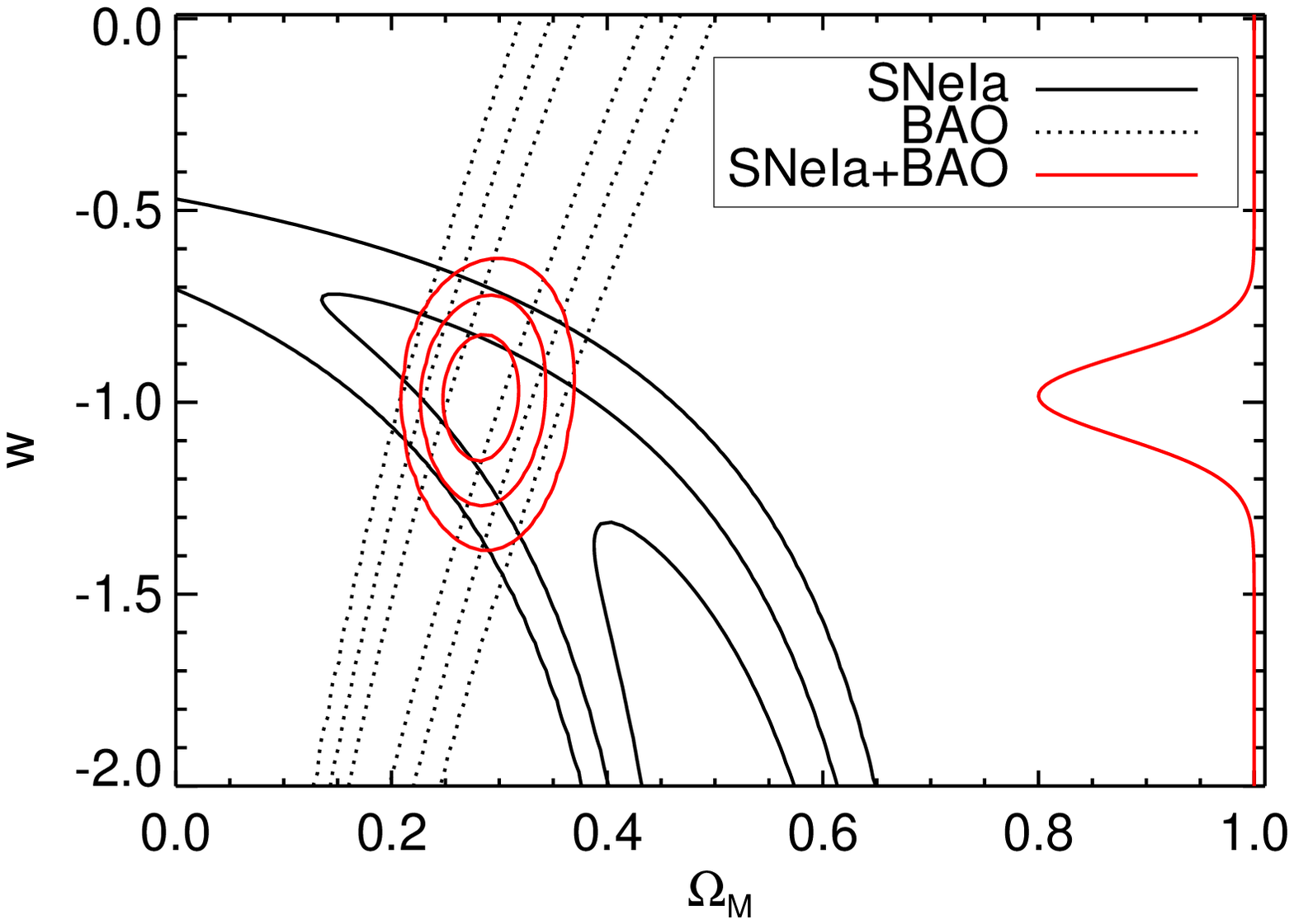}
\caption{The \OM-$w$ 1$\sigma$, 2$\sigma$, and 3$\sigma$ contours from the ESSENCE + nearby sample for MLCS2k2 with the  ``glosz'' $A_V$ prior and with the SALT fitter.  The baryon acoustic oscillation (BAO) constraints are from \citet{eisenstein05}.}
\label{fig:mlcs_prior_glosz_salt_om_w}
\end{figure}

\begin{figure}
\plottwo{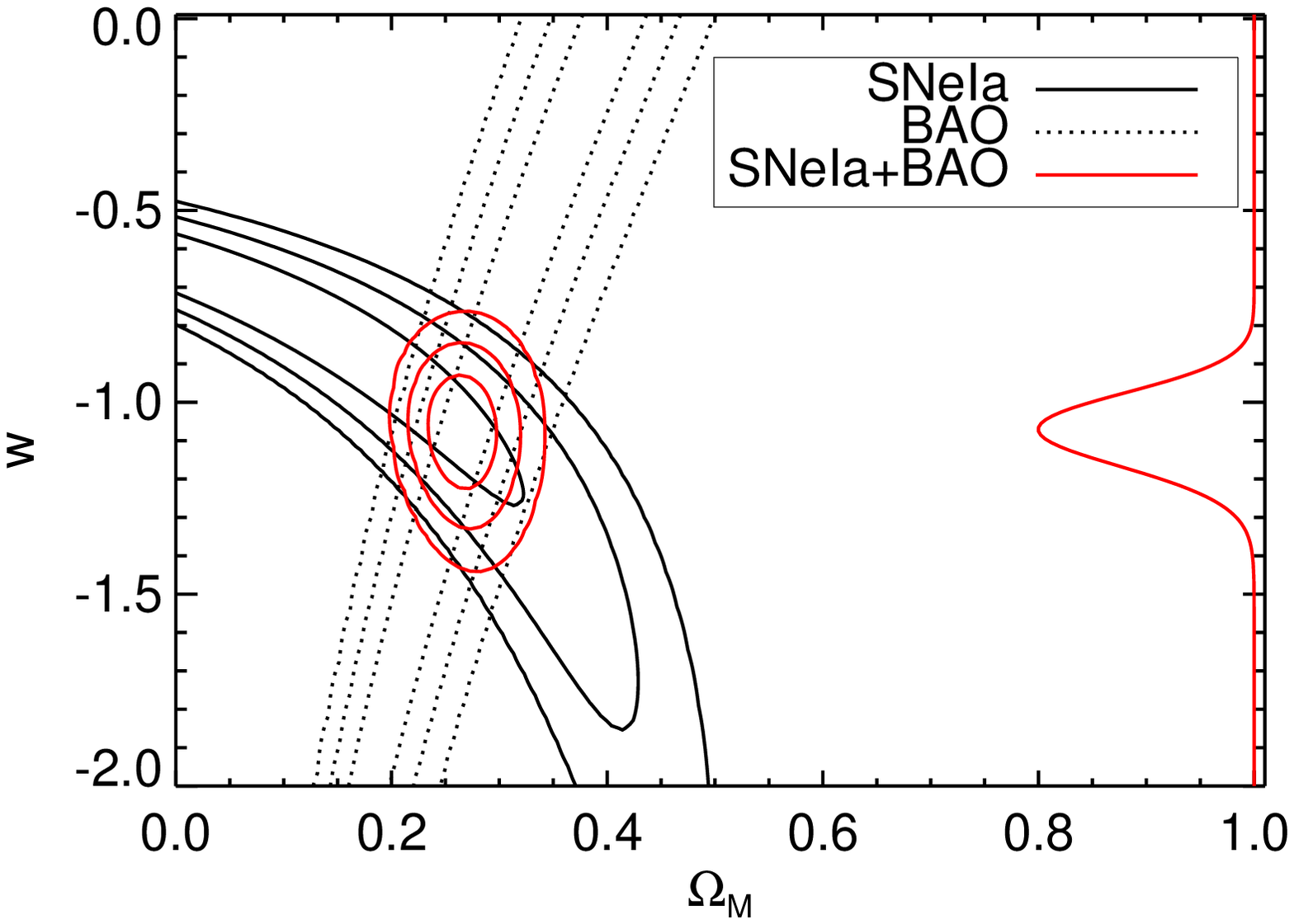}{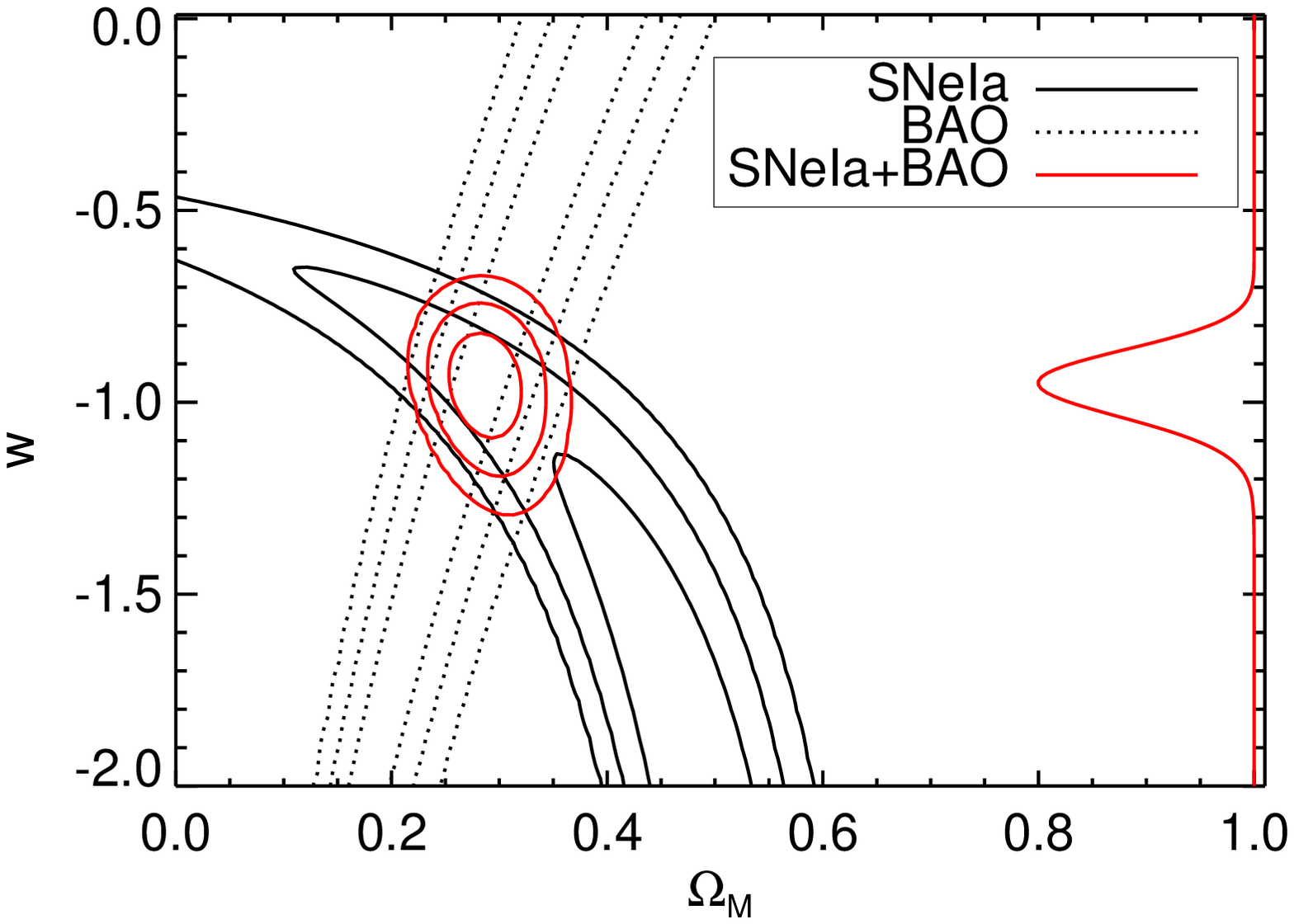}
\caption{The \OM-$w$ contours from the SNLS + ESSENCE + nearby sample for MLCS2k2 with ``glosz'' $A_V$ prior and for the SALT fitter.
  The baryon acoustic oscillation (BAO) constraints are from \citet{eisenstein05}.}
\label{fig:joint_mlcs_prior_glosz_salt_om_w}
\end{figure}

\begin{figure}
\plottwo{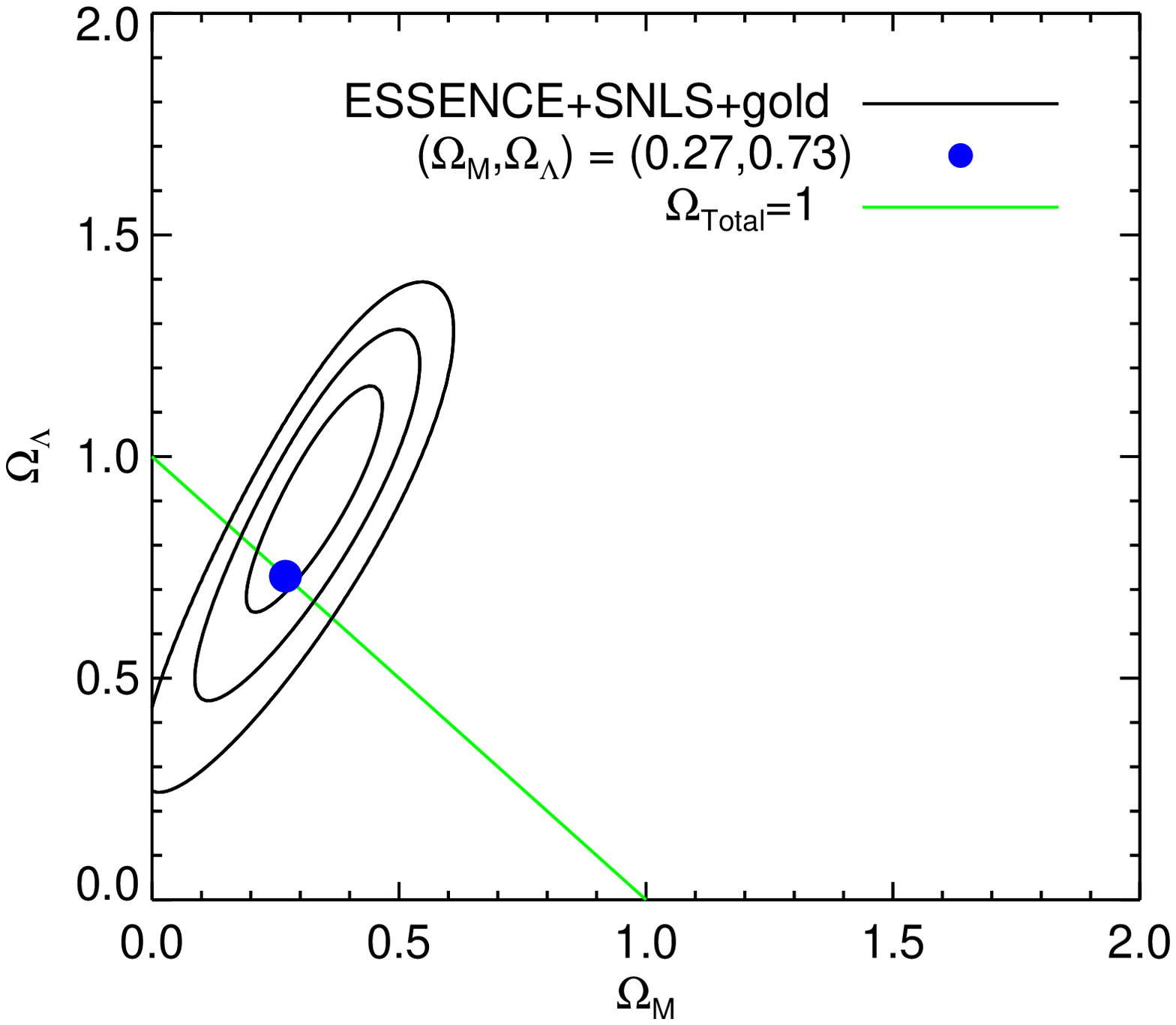}{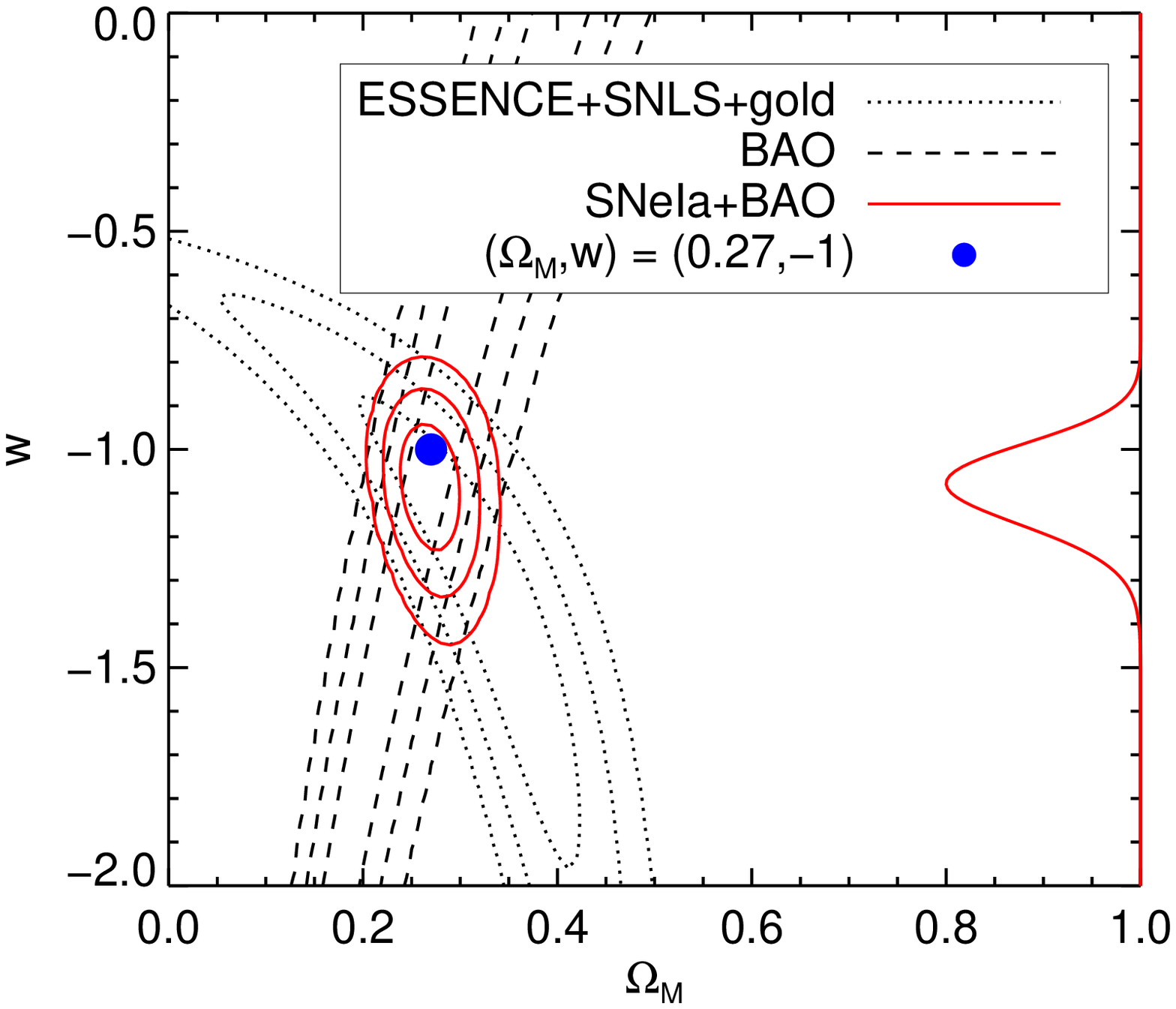}
\caption{The \snia\ (\OM, \OL) and (\OM, $w$) contours from combining 
the MLCS2k2 luminosity distances
for the ESSENCE 
\sneia\ analyzed here in combination with the nearby \sneia, SNLS
\sneia, and the Riess ``gold'' sample. 
The diagonal line in the (\OM, \OL) plot represents a flat Universe, \OT$=$\OM$+$\OL$=1$.  
From the \sneia\ data alone, an empty Universe is ruled out at
$4.5$~$\sigma$, an (\OM, \OL) = $(0.3, 0)$ Universe at $10$~$\sigma$,
and an (\OM, \OL) = $(1, 0)$~$\sigma$ Universe at $>20$~$\sigma$.
The best combination of data will come after a complete analysis of
the calibration and systematic errors of all the data sets.  We offer
this interim result to indicate the potential of combining low-z,
ESSENCE, and supernovae at redshifts beyond $1$.
}
\label{fig:OMOL_OMw_joint_riess04}
\end{figure}

\begin{figure}
\plotone{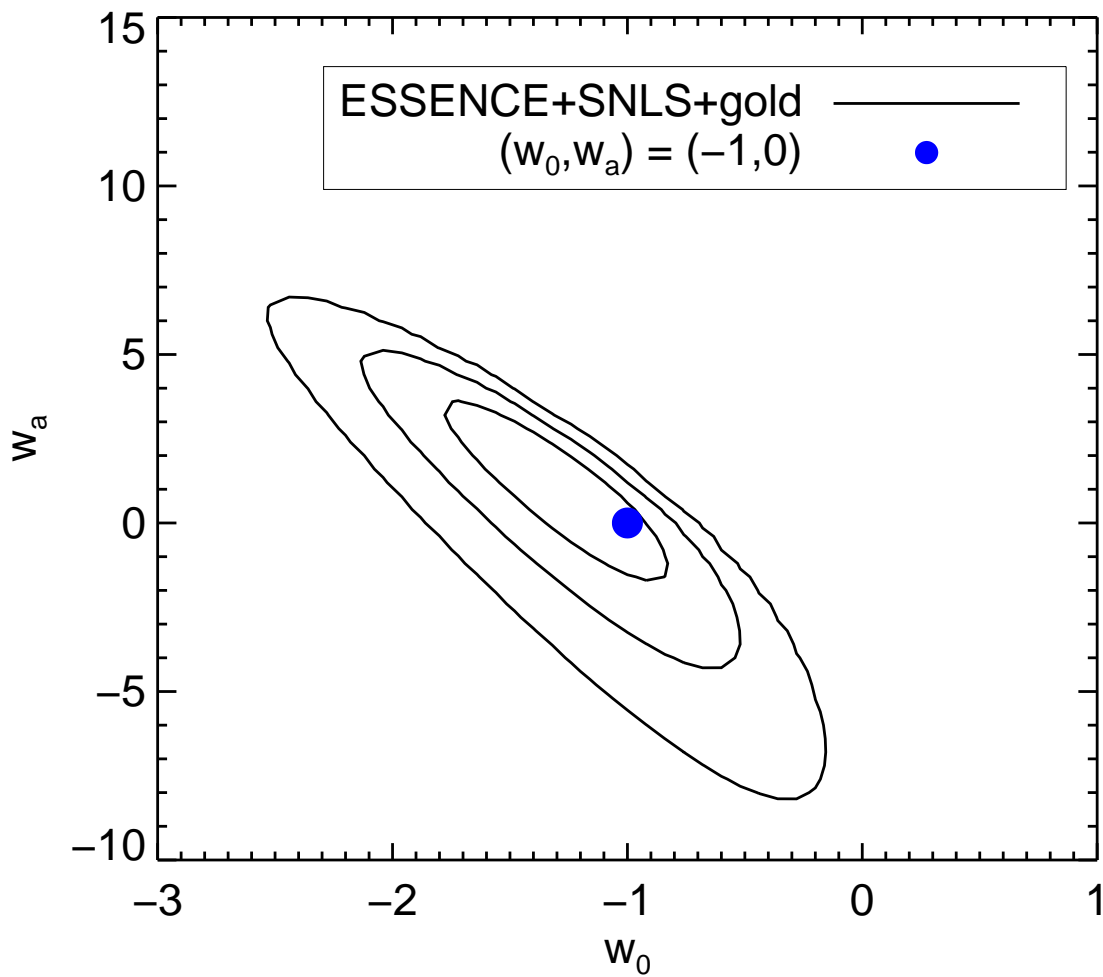}
\caption{Combined constraints on ($w_0$, $w_a$) using the MLCS2k2 luminosity distances
for the ESSENCE 
\sneia\ analyzed here in combination with the nearby \sneia, SNLS
\sneia, and the Riess ``gold'' sample.  
Here we are considering a two-parameter representation of the dark energy equation-of-state parameter, $w=w_0+w_a (1-a)$.
Instead of the BAO constraints
we have simply taken \OM$=0.27\pm0.03$.
(See cautionary note from Fig.~\ref{fig:OMOL_OMw_joint_riess04}.)
}
\label{fig:w0wa_joint_riess04}
\end{figure}

\pagebreak
\begin{deluxetable}{ll}
\tablewidth{0pc}
\tablecaption{MLCS2k2 Fit Parameter Quality Cuts}
\tablehead{
\colhead{Fit Parameter} & \colhead{Requirement} \\
}
\startdata
$\chi^2_\nu$                       &  $\chi^2_\nu \le 3$                    \\
\# Degrees of Freedom              &  ${\rm DoF} \ge 4$                      \\
$\Delta$                           &  $-0.4 \le \Delta \le 1.7$                \\
Time of Maximum uncertainty        & ${\rm Tmax}_{\rm err} \le 2.0$ rest-frame days \\
First observation w/ SNR $> 5$     & $\le +4$ days                          \\
Last observation  w/ SNR $> 5$     & $\ge +9$ days                         \\
\enddata
\tablecomments{See Tables.~\ref{tab:mlcs_fits_prior_glosz} for the MLCS2k2 fit parameters used for the cosmological analysis presented in this paper. These selection criteria were derived based on Monte Carlo simulations discussed in \S\ref{sec:simulation}.  The number of degrees of freedom is the number of light-curve points with SNR$>5$ minus the 4 independent MLCS2k2 fit parameters: $m_V$, $\Delta$, $A_V$ and $T_{\rm max}$.
}
\label{tab:mlcs_cuts}
\end{deluxetable}

\begin{deluxetable}{ccc|cc}
\centering
\tablewidth{0pt}
\tablecaption{``glosz'' Window Function Parameters}
\tablehead{\colhead{$z$} & \colhead{$A_{1/2}$} &
\colhead{$\sigma_A$} & \colhead{$\Delta_{1/2}$} &
\colhead{$\sigma_\Delta$} }
\startdata
0.35 & 1.35 & 2.2 & 0.93 & 2.4 \\
0.40 & 1.05 & 2.5 & 0.75 & 2.4 \\
0.45 & 0.88 & 2.6 & 0.60 & 2.6 \\
0.50 & 0.67 & 2.8 & 0.43 & 2.6 \\
0.55 & 0.48 & 3.5 & 0.29 & 2.7 \\
0.60 & 0.33 & 4.0 & 0.17 & 2.8 \\
0.65 & 0.20 & 5.0 & 0.05 & 3.0 \\
0.70 & 0.10 & 6.0 & $-0.09$ & 3.3 \\
0.75 & 0.05 & 7.5 & $-0.25$ & 3.3 \\
\enddata
\label{tab:gloszwindow}
\end{deluxetable}

\begin{deluxetable}{ll}
\tablewidth{0pc}
\tablecaption{SALT Fit Parameter Quality Cuts}
\tablehead{
\colhead{Fit Parameter} & \colhead{Requirement} \\
}
\startdata
$\chi^2_\nu$                       &  $\chi^2_\nu \le 3$                    \\
\# Degrees of Freedom              &  ${\rm DoF} \ge 5$                      \\
stretch                            &  $0.5 \le s < 1.4$                \\
Time of Maximum uncertainty        & ${\rm Tmax}_{\rm err} \le 2.0$ rest-frame days \\
\# Observations after B-band maximum & $> 1$  \\
First observation w/ SNR $> 5$     & $\le +5$ days                          \\
\enddata
\tablecomments{See Tables.~\ref{tab:salt_fits} for the SALT fit parameters used for the cosmological analysis presented in this paper.  These selection criteria were based on \citet{astier06} with additional sanity checks on the stretch parameter and uncertainty in the time of maximum light.
The number of degrees of freedom is the number of light-curve points with SNR$>5$ minus the 4 independent SALT fit parameters: $m_B$, strech, color, and $T_{\rm max}$.
}
\label{tab:salt_cuts}
\end{deluxetable}

\begin{deluxetable}{ll}
\tablewidth{0pc}
\tablecaption{Sources of Increased $\mu$ Dispersion}
\tablehead{
\colhead{Source} & \colhead{$\sigma_\mu$}  \\
}
\startdata
Flatfielding              &  0.01  \\
Focal-plane PSF           &  0.02  \\
Field-field zeropoint     &  0.01  \\
Image subtraction         &  0.01  \\
\tableline
Subtotal (quadrature sum) &  0.026 \\
\tableline
Gravitational lensing     &  0.03  \\
\tableline
Total (quadrature sum)    &  0.04  \\
\enddata
\tablecomments{
The photometric and astrophysical uncertainties that add increased
scatter, but no bias, to the measured \snia distance moduli, $\mu$.
}
\label{tab:photscatter}
\end{deluxetable}

\begin{deluxetable}{lllll}
\tabletypesize{\small}
\rotate
\tablewidth{0pc}
\tablecaption{Potential Sources of Systematic Error on the Measurement of $w$}
\tablehead{
\colhead{Source}                      & \colhead{$dw/dx$}   & \colhead{$\Delta x$} &   $\Delta_w$   & Notes \\
}
\startdata
Phot. errors from astrometric uncertainties of faint objects & 1/mag & 0.005 mag   &      0.005     &     \\ 
Bias in diff im photometry            &      0.5  / mag     &       0.002 mag      &      0.001     &     \\
CCD linearity                         &      1    / mag     &       0.005 mag      &      0.005     &     \\
Photometric zeropoint diff in $R$,$I$ &      2    / mag     &       0.02  mag      &      0.04      &  \\ 
Zpt. offset between low and high z    &      1    / mag     &       0.02  mag      &      0.02      &  \\
K-corrections                         &      0.5  / mag     &       0.01  mag      &      0.005     &  \\
Filter passband structure             &      0    / mag     &       0.001 mag      &      0         &     \\
Galactic extinction                   &      1    / mag     &       0.01  mag      &      0.01      &  \\
Host galaxy $R_V$                     &      0.02 / $R_V$   &       0.5            &      0.01      &  ``glosz'' \\
Host galaxy extinction treatment      &     0.08            &  prior choice        &      0.08      &  different priors \\ 
Intrinsic color of \sneia             &       3   / mag     &       0.02  mag      &      0.06      &  interacts strongly with prior \\

Malmquist bias/selection effects      &      0.7  / mag     &       0.03  mag      &      0.02      &  ``glosz'' \\

SN~Ia evolution                       &      1    / mag     &       0.02  mag      &      0.02      &  \\
Hubble bubble                         & 3/$\delta H_{\rm effective}$ & 0.02        &      0.06      &  \\
Gravitational lensing                 & 1/$\sqrt{N}$ / mag  &       0.01  mag      &  $< 0.001$     &  \citet{holz05}  \\
Grey dust                             &      1    / mag     &       0.01  mag      &      0.01      &   \\

Subtotal w/o extinction+color         &       \nodata       &    \nodata           &      0.082     & \\
Total                                 &       \nodata       &    \nodata           &      0.13      & \\
\tableline
Joint ESSENCE+SNLS comparison         &       \nodata       &    \nodata           &      0.02      & photometric system \\
Joint ESSENCE + SNLS Total            &       \nodata       &    \nodata           &      0.13      & \\

\enddata
\tablecomments{The systematic error table for this first ESSENCE
cosmological analysis.  The issue of treatment of $A_V$ and color
distribution is clearly the dominant systematic effect and will need to be
seriously addressed to reduce our systematic errors to our target of 5\%.
}
\label{tab:systematics}
\end{deluxetable}

\begin{deluxetable}{lrrrr}
\tablewidth{0pc}
\tablecaption{Effect of different fixed $R_V$ on $w$ for the two different $A_V$ priors considered in our MLCS2k2 analysis of the ESSENCE+nearby sample.}
\tablehead{ 
\colhead{$A_V$ prior} & \colhead{$R_V$ value} & \colhead{$w$} & \colhead{$w$-$w_{R_V=3.1}$} \\
\colhead{} & \colhead{} & \colhead{} & \colhead{for given prior} \\
}
\startdata
glosz                &  2.1 & $-0.986_{-0.114}^{+0.116}$ & $+0.061$ \\
glosz                &  3.1 & $-1.047_{-0.124}^{+0.125}$ &    \nodata \\
glosz                &  4.1 & $-1.073_{-0.120}^{+0.121}$ & $-0.026$ \\
glos                 &  2.1 & $-0.932_{-0.114}^{+0.116}$ & $+0.025$ \\
glos                 &  3.1 & $-0.957_{-0.124}^{+0.127}$ &    \nodata \\
glos                 &  4.1 & $-1.039_{-0.131}^{+0.134}$ & $-0.082$ \\
exponential          &  2.1 & $-0.855_{-0.122}^{+0.126}$ & $+0.027$ \\
exponential          &  3.1 & $-0.882_{-0.130}^{+0.134}$ &    \nodata \\
exponential          &  4.1 & $-0.808_{-0.141}^{+0.147}$ & $+0.074$ \\
\enddata
\tablecomments{Systematic effect of choosing different fixed $R_V$ values for the different $A_V$ priors discussed here for MLCS2k2.}
\label{tab:w_rv_sys}
\end{deluxetable}
\clearpage

\begin{deluxetable}{lrrrrrr}
\tablewidth{0pc}
\tablecaption{Cosmological Parameters from ESSENCE+nearby and BAO Constraints}
\tablehead{
\colhead{}       & \colhead{\#\sneia{\tablenotemark{a}}} & \colhead{\lcdm{\tablenotemark{b}}} & \multicolumn{3}{c}{flat, constant-$w$ (marg. 1D)} \\
\colhead{Sample} & & \colhead{\chisqnu}  & \colhead{\chisqnu} & \colhead{$w_0$} & \colhead{\OM} \\
}
\startdata
{\bf MLCS2k2: ``glosz'' } & & & & & \\
 All ESSENCE+nearby                 &    102 &   0.96 &   0.96 & $-1.047_{-0.124}^{+0.125}$ & $ 0.274_{-0.020}^{+0.032}$ \\
  ESSENCE only                      &     57 &   0.88 &   0.91 & \nodata & \nodata \\
   nearby only                      &     45 &   1.00 &   1.01 & \nodata & \nodata \\
{\bf SALT}                         & & & & & \\
 All ESSENCE+nearby                 &    106 &   2.62 &   2.66 &  $-0.988_{-0.109}^{+0.110}$ & $ 0.284_{-0.020}^{+0.031}$ \\
  ESSENCE only                      &     60 &   4.64 &   4.72 & \nodata & \nodata \\
   nearby only                      &     46 &   1.01 &   1.04 & \nodata & \nodata \\
\enddata
\tablenotetext{a}{$0.015<z$.}
\tablenotetext{b}{\lcdm\ refers to a universe with ($w$, \OM, \OL) = $(-1, 0.27, 0.73)$.}
\tablecomments{
The ESSENCE cosmological results given here are for our favored MLCS2k2 ``glosz'' $A_V$ prior and the SALT fitter of \citet{guy05}.
The DoF for the \lcdm\ model is the number of
\sneia\ in each set minus the one fit parameters, \scriptM.
The DoF for the best-fit model is the number of \sneia\ minus
the three fit parameters ($w$, \OM, \scriptM).
For the subsets, the same cosmological fit is used 
but \scriptM\ is allowed to float.
The \chisqnu\ for \lcdm\ for the nearby set is $1$ by construction.
The appropriate additional $\sigma_\mu$ to add in quadrature 
to recover the full intrinsic dispersion of \sneia{}
is determined by requiring \chisqnu\ of the nearby set to be $1$ for \lcdm{}
with an assumed peculiar velocity of $400$~km/s.
The value for $w_0$ is marginalized over \OM\ assuming a flat,
\OMLflat\ Universe.
Note that the \chisqnu\ values for the marginalized 1D values
are higher than the \lcdm\ model.  This is possible because the
mean marginalized 1D values are not the points of lowest \chisq.
This indicates that there is no reason to favor the
marginalized 1D values over the \lcdm\ model.
}
\label{tab:results}
\end{deluxetable}

\begin{deluxetable}{lrrrrrr}
\tablewidth{0pc}
\tablecaption{Joint Cosmological Parameters from ESSENCE+SNLS+nearby and BAO Constraints}
\tablehead{
\colhead{}       & \colhead{\#\sneia{\tablenotemark{a}}} & \colhead{\lcdm{\tablenotemark{b}}} & \multicolumn{3}{c}{flat, constant-$w$ (marg. 1D)} \\
\colhead{Sample} & & \colhead{\chisqnu} & \colhead{\chisqnu} & \colhead{$w_0$} & \colhead{\OM} \\
}
\startdata
{\bf MLCS2k2: ``glosz'' } & & & & & \\
 All ESSENCE+SNLS+nearby            &    162 &   0.90 &   0.91  & $-1.069_{-0.093}^{+0.091}$ & $ 0.267_{-0.018}^{+0.028}$ \\
   ESSENCE only                     &     60 &   0.91 &   0.93  & \nodata & \nodata \\
   SNLS only                        &     57 &   0.82 &   0.82  & \nodata & \nodata \\
   nearby only                      &     45 &   0.99 &   1.01  & \nodata & \nodata \\
{\bf SALT}                         & & & & & \\
 All ESSENCE+SNLS+nearby            &    178 &   2.76 &   2.79  & $-0.958_{-0.090}^{+0.088}$ & $ 0.288_{-0.019}^{+0.029}$ \\
   ESSENCE only                     &     64 &   4.77 &   4.78  & \nodata & \nodata \\
   SNLS only                        &     68 &   2.07 &   2.12  & \nodata & \nodata \\
   nearby only                      &     46 &   0.99 &   1.02  & \nodata & \nodata \\
\enddata
\tablenotetext{a}{$0.015<z$.}
\tablenotetext{b}{\lcdm\ refers to a universe with ($w$, \OM, \OL) = $(-1, 0.27, 0.73)$.}
\tablecomments{
See notes for Table~\ref{tab:results}.
We include the full sample here without the redshift cut of the
ESSENCE-only analysis to consistently include all of the usable
\sneia.
}
\label{tab:joint_results}
\end{deluxetable}



\begin{deluxetable}{lrrrrrrrrrrrrrrl}
\rotate
\tabletypesize{\scriptsize}
\tablewidth{0pc}
\tablecaption{MLCS2k2 ``glosz'' Luminosity Distances of all \sneia.}
\tablehead{
\colhead{Name} & \colhead{$z$} & \colhead{$\sigma_z${\tablenotemark{a}}} & \colhead{$\mu$} & \colhead{$\sigma_\mu${\tablenotemark{b}}} & \colhead{$A_V$}  & \colhead{$\sigma_{A_V}$} & \colhead{$\Delta$} & \colhead{$\sigma_\Delta$} & \colhead{$T_{\rm max}$} & \colhead{$\sigma_{T_{\rm max}}$} & \colhead{\chisqnu} & \colhead{DoF} & \colhead{$T_{\rm first}$} & \colhead{$T_{\rm last}$} & \colhead{Failed} \\
 &  &  & [mag] & [mag] & [mag] & [mag] & & & [MJD] & [MJD] & & & [day] & [day] & \\
}
\startdata
b010       & $ 0.5910$ & $  0.007$ & $  42.984$ & $  0.199$ & $  0.104$ & $  0.094$ & $  -0.166$ & $  0.182$ & $  52592.80$ & $   2.69$ & $   1.01$ & $    3$ & $   0.22$ & $  19.70$ & x \\
b013       & $ 0.4260$ & $  0.004$ & $  41.976$ & $  0.205$ & $  0.170$ & $  0.149$ & $  +0.034$ & $  0.141$ & $  52586.33$ & $   1.51$ & $   0.22$ & $   12$ & $  -2.21$ & $  35.64$ &  \\
b016       & $ 0.3290$ & $  0.003$ & $  41.349$ & $  0.411$ & $  0.359$ & $  0.275$ & $  +0.190$ & $  0.384$ & $  52587.76$ & $   2.29$ & $   3.31$ & $    1$ & $  -0.56$ & $  16.76$ & x \\
b020       & $ 0.4250$ & $  0.003$ & $  41.766$ & $  0.371$ & $  0.202$ & $  0.204$ & $  +0.059$ & $  0.323$ & $  52599.88$ & $   0.90$ &   \nodata & $    0$ & $  -8.94$ & $   7.21$ & x \\
d033       & $ 0.5310$ & $  0.008$ & $  42.960$ & $  0.138$ & $  0.085$ & $  0.091$ & $  -0.322$ & $  0.109$ & $  52934.26$ & $   1.87$ & $   0.98$ & $    5$ & $  -2.12$ & $  23.41$ &  \\
d058       & $ 0.5830$ & $  0.009$ & $  43.103$ & $  0.151$ & $  0.119$ & $  0.104$ & $  -0.470$ & $  0.077$ & $  52941.76$ & $   1.53$ & $   1.70$ & $    7$ & $  -6.11$ & $  33.04$ & x \\
d083       & $ 0.3330$ & $  0.002$ & $  40.709$ & $  0.104$ & $  0.084$ & $  0.080$ & $  -0.273$ & $  0.061$ & $  52936.71$ & $   0.83$ & $   0.25$ & $   20$ & $  -4.17$ & $  43.08$ &  \\
d084       & $ 0.5190$ & $  0.007$ & $  42.948$ & $  0.274$ & $  0.221$ & $  0.188$ & $  -0.197$ & $  0.207$ & $  52937.24$ & $   1.81$ & $   1.15$ & $    4$ & $  -4.01$ & $  13.77$ &  \\
\enddata
\tablenotetext{a}{We add a 400~km/s peculiar velocity dispersion in quadrature to these redshift uncertainties for our cosmological fits.}
\tablenotetext{b}{An ``intrinsic'' dispersion of $0.10$ should be added in quadrature to these values output by MLCS2k2~\citep{jha06b} to fully account for the intrinsic dispersion of \sneia.}
\tablecomments{
The luminosity distances and extinctions as determined by
MLCS2k2 of the full ESSENCE \sneia and nearby sample using the ``glosz'' prior~\cite{hatano98,commins04,riello05} described in \ref{sec:distances}.
\snia fit marked as ``Failed'' did not pass the MLCS2k2 quality cuts given Table~\ref{tab:mlcs_cuts}.
See Table.~\ref{tab:mlcs_cuts} and \S\ref{sec:distances} for further discussion of the quality cuts applied here.
Redshifts of \sneia come from SNID fits \citep{miknaitis07,blondin07}.}
\label{tab:mlcs_fits_prior_glosz}
\end{deluxetable}



\begin{deluxetable}{lrrrrrrrrrrrrrrl}
\rotate
\tabletypesize{\scriptsize}
\tablewidth{0pc}
\tablecaption{SALT Luminosity Distances of all \sneia.}
\tablehead{
\colhead{Name} & \colhead{$z$} & \colhead{$\sigma_z${\tablenotemark{a}}} & \colhead{$\mu$} & \colhead{$\sigma_\mu${\tablenotemark{b}}} & \colhead{color}  & \colhead{$\sigma_{\rm color}$} & \colhead{stretch} & \colhead{$\sigma_{\rm stretch}$} & \colhead{$T_{\rm max}$} & \colhead{$\sigma_{T_{\rm max}}$} & \colhead{\chisqnu} & \colhead{DoF} & \colhead{Rise} & \colhead{Tail} & \colhead{Failed} \\
 & & & [mag] &  [mag] & & & & & [MJD] & [MJD] & & & & & \\
}
\startdata
b010       & $ 0.5910$ & $  0.007$ & $  43.524$ & $  0.241$ & $ -0.127$ & $  0.086$ & $   1.232$ & $  0.141$ & $  52589.42$ & $   4.98$ & $   0.99$ & $   15$ &       0 &      19 & x \\
b013       & $ 0.4260$ & $  0.004$ & $  42.144$ & $  0.085$ & $ +0.077$ & $  0.039$ & $   1.012$ & $  0.050$ & $  52586.02$ & $   0.94$ & $   0.71$ & $   13$ &       2 &      15 &  \\
b016       & $ 0.3290$ & $  0.003$ & $  41.876$ & $  0.169$ & $ +0.246$ & $  0.102$ & $   1.121$ & $  0.002$ & $  52579.73$ & $   0.05$ & $   0.73$ & $    6$ &       0 &      10 &  \\
b020       & $ 0.4250$ & $  0.003$ & $  41.696$ & $  0.315$ & $ +0.068$ & $  0.176$ & $   0.840$ & $  0.111$ & $  52599.81$ & $   1.07$ &  \nodata  & $    0$ &       1 &       3 & x \\
d033       & $ 0.5310$ & $  0.008$ & $  43.271$ & $  0.136$ & $ -0.156$ & $  0.059$ & $   1.150$ & $  0.088$ & $  52932.50$ & $   1.70$ & $   1.74$ & $   17$ &       2 &      19 &  \\
d058       & $ 0.5830$ & $  0.009$ & $  43.011$ & $  0.092$ & $ +0.188$ & $  0.068$ & $   1.095$ & $  0.006$ & $  52938.34$ & $   0.09$ & $   4.24$ & $   16$ &       3 &      17 &  \\
d083       & $ 0.3330$ & $  0.002$ & $  40.908$ & $  0.028$ & $ -0.020$ & $  0.014$ & $   1.180$ & $  0.015$ & $  52936.99$ & $   0.18$ & $   5.44$ & $   15$ &       4 &      15 &  \\
d084       & $ 0.5190$ & $  0.007$ & $  43.385$ & $  0.187$ & $ -0.015$ & $  0.090$ & $   1.095$ & $  0.114$ & $  52935.42$ & $   2.17$ & $   2.30$ & $   17$ &       4 &      17 &  \\
\enddata
\tablenotetext{a}{We add a 400~km/s peculiar velocity dispersion in quadrature to these redshift uncertainties for our cosmological fits.}
\tablenotetext{b}{An ``intrinsic'' dispersion of $\sigma_\mu=0.13$ should be added in quadrature to these values output by SALT~\citep{guy05} 
to fully account for the intrinsic dispersion of \sneia.}
\tablecomments{
The luminosity distances and extinctions as determined by
SALT of the full SNLS, ESSENCE, and nearby \snia sample.
\snia fit marked as ``Failed'' did not pass the SALT quality cuts given Table~\ref{tab:salt_cuts}.
The SALT fitter quotes minimum chi-sq values for the lightcurve fit
parameters rather than estimated mean estimated parameters.
See Table.~\ref{tab:salt_cuts} and for the quality cuts applied here.
}
\label{tab:salt_fits}
\end{deluxetable}

\end{document}